\documentclass[aps,prl,reprint,twocolumn,amsmath,superscriptaddress,amssymb,showpacs]{revtex4-1}
\usepackage{amsmath,braket,amssymb}
\usepackage[final]{graphicx}
\usepackage{subfigure}
\usepackage{xcolor}
\usepackage[english]{babel}
\usepackage[utf8]{inputenc}
\usepackage{color}
\usepackage{mathtools}

\usepackage[colorlinks,citecolor=darkBlue,linkcolor=darkBlue,
urlcolor=blue,hyperindex]{hyperref}

\usepackage[normalem]{ulem}
\normalfont

\definecolor{forestgreen}{rgb}{0.08, 0.4, 0.13}
\definecolor{darkBlue}{rgb}{0.08, 0.13, 0.4}

\newcommand{\dia}[3]{\raisebox{#2pt}{\includegraphics[height=#3pt]{#1}}}

\begin{document}

\title{Universal behavior beyond multifractality of wave-functions at measurement--induced phase transitions}

\author{Piotr Sierant}
\affiliation{ICTP - The Abdus Salam International Center for Theoretical Physics, Strada Costiera 11, 34151 Trieste, Italy}
 \affiliation{Institute of Theoretical Physics, Jagiellonian University in Krakow, \L{}ojasiewicza 11, 30-348 Krak\'ow, Poland }
\author{Xhek Turkeshi}
\affiliation{ICTP - The Abdus Salam International Center for Theoretical Physics, Strada Costiera 11, 34151 Trieste, Italy}
\affiliation{SISSA - International School of Advanced Studies, via Bonomea 265, 34136 Trieste, Italy}
\date\today

\begin{abstract}
	We investigate the structure of many-body wave functions of 1D quantum circuits with local measurements employing the participation entropies. The leading term in system size dependence of participation entropy indicates a model dependent multifractal scaling of the wave-functions at any non-zero measurement rate. The sub-leading term contains universal information about measurement-induced phase transitions and plays the role of an order parameter, being constant non-zero in the error correcting phase and vanishing in the quantum Zeno phase.
	We provide robust numerical evidence investigating a variety of quantum many-body systems, and provide an analytical interpretation of this behavior expressing the participation entropy in terms of partition functions of classical statistical models in 2D.
\end{abstract}

\maketitle

The evolution of quantum systems driven by the interplay of unitary dynamics and the monitoring action of an environment follows a quantum trajectory through the Hilbert space~\cite{carmichael2009open, Dalibard92,Molmer93, breuer2002theory,gardiner2004quantum,wiseman2009quantum,daley2014quantum,basche1995direct,gleyzes2007quantum,vijay2011observation,robledo2011highfidelity,minev2019to}.
While the former generates coherences and spreads information throughout the system, the latter partially resolves the state, which is collapsed according to the Born-Von Neumann postulate. 
In the many-body framework, this competition gives rise to 
dynamical phases captured by non-linear functions of the density matrix
separated by so-called measurement-induced phase transitions
(\textbf{MIPTs})~\cite{skinner2019measurementinduced,chan2019unitaryprojective,nahum2021measurement,jian2021yanglee,shtanko2020classical,zabalo2020critical,szyniszewski2019entanglement,snizhko2020quantum,jian2020criticality,sang2020entanglement,shi2020entanglement,ippoliti2021postselectionfree,fuji2020measurementinduced,rossini2020measurementinduced,lunt2020measurementinduced,tang2021quantum,gopalakrishnan2021entanglement,buchhold2021effective,medina2021entanglement,lu2021entanglement,lunt2021dimensional,li2020conformal,turkeshi2020measurementinduced,lavasani2021topological,sang2021measurementprotected,lavasani2021measurementinduced,block2021measurementinduced,czischek2021simulating,zhang2020nonuniversal,turkeshi2021measurementinduced,chen2020emergent,noel2021observation,ippoliti2021fractal,agrawal2021entanglement,boorman2021diagnosing,tang2020measurementinduced, goto2020measurementinduced,Dhar16, lang2020entanglement, biella2021manybody, turkeshi2021measurementinduced_2, Botzung21, sierant2021dissipative,cao2019entanglement, muller2021measurementinduced, alberton2021entanglement, minato2021fate, Zhang21}.

Random quantum circuits, believed to model generic chaotic quantum dynamics \cite{Nahum17, Rakovszky18diffusive, Keyserlingk18operator, Khemani18, Chan18, Nahum18, Chang19}, become minimal models to investigate MIPTs when their discrete space-time is punctured by local projective measurements~\cite{li2018quantum,li2019measurementdriven}. 
Entanglement provides natural measures to explore the phase diagram of these models~\cite{gullans2020scalable}. 
Their scaling 
\begin{figure}[h!]
	\centering
	\includegraphics[width=\columnwidth]{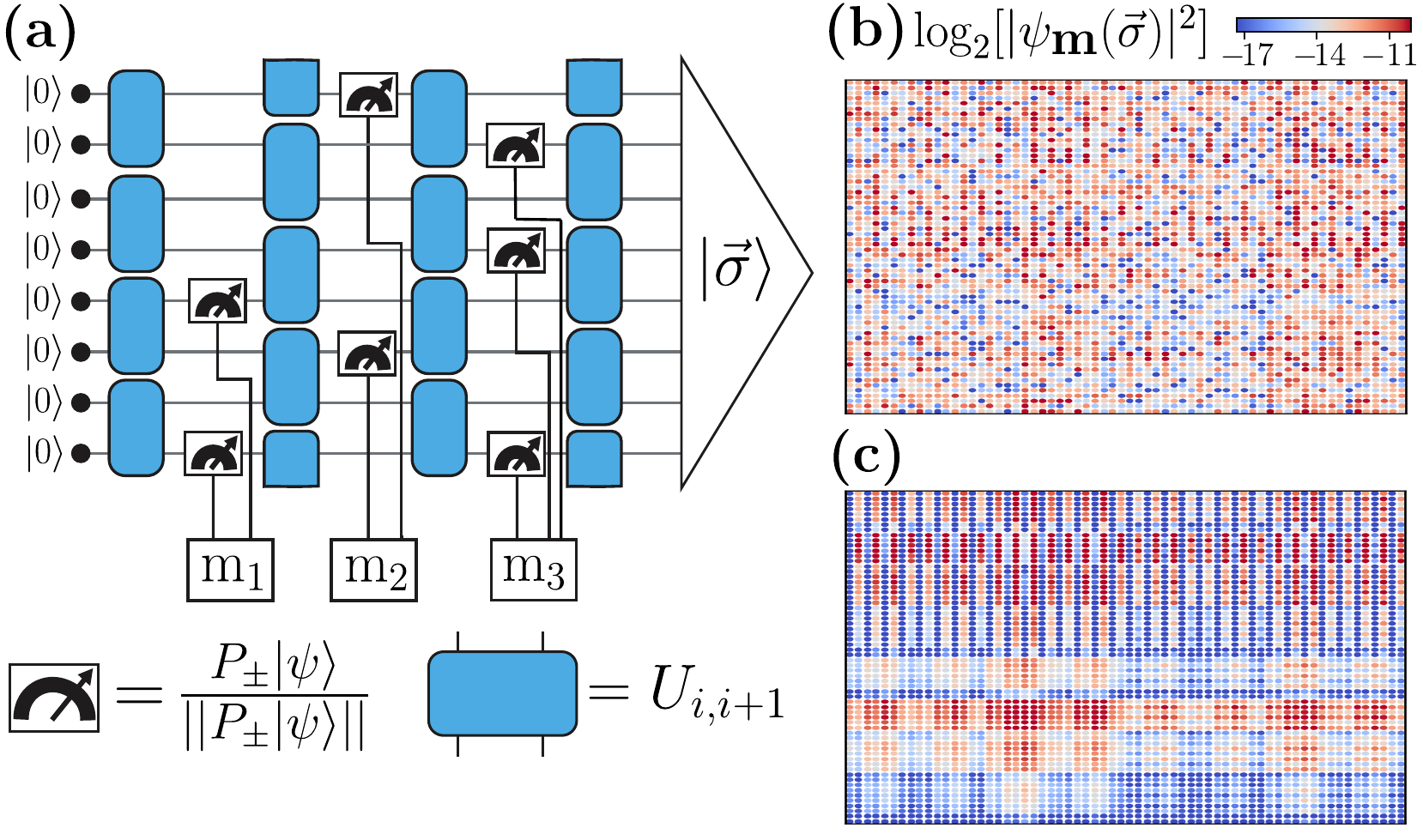}
	\caption{\label{fig1} \textbf{(a)}: Quantum circuit consisting of random unitary 2-qubit gates $U_{i,i+1}$ and local measurements $P_{\pm}=(1\pm Z)/2$.
	\textbf{(b)} and \textbf{(c)}: $\log_2[|\psi_\mathbf{m}(\vec{\sigma})  | ^2]$, where  $\psi_\mathbf{m}(\vec{\sigma}) = \braket{ \vec{\sigma} | \psi_{\mathrm{\textbf{m}}} }$ is the many-body wave function, 
	which changes between volume-law phase [$p=0.1$, \textbf{(b)}] and area-law phase [$p=0.4$, \textbf{(c)}]. We analyze this change using the PEs~\eqref{sqdef}. The data is for a Haar circuit of $L=12$ qubits.} 
	\vspace{-0.7cm}
\end{figure}
behavior with system size captures key facets of the dynamical phases and justifies their intriguing interpretation in terms of quantum error correction~\cite{choi2020quantum, li2021statistical,fidkowski2021how,gullans2020dynamical,ippoliti2021entanglement,bao2021symmetry}.
Specifically, when the measurement rate is tuned, the system undergoes a transition from an error-correcting phase to a quantum Zeno phase, characterized respectively by an extensive (``volume law'') and a sub-extensive (``area law'') scaling of the entanglement entropy. 

The local measurements partially project the state $\ket{\psi_\mathbf{m}}$ of the circuit affecting the structure of many-body wave-function $\psi_\mathbf{m}(\vec{\sigma}) = \braket{ \vec{\sigma} | \psi_{\mathrm{\textbf{m}}} }$, at the quantum trajectory $\mathbf{m}$, in the basis $\{\ket{\vec{\sigma}}\}$ (Fig.~\ref{fig1}). The structural changes are probed via the participation entropies (\textbf{PEs})
\begin{align}
	S_q = \frac{1}{1-q} \log_2 \sum_{{\vec{\sigma}}} |\psi_\mathbf{m}(\vec{\sigma})|^{2q} \equiv D_q L + c_q,
	\label{sqdef}
\end{align}
where ${q>0}$ and the second equality parametrizes the scaling of $S_q$ with system size $L$ by a fractal dimension $D_q$ and a sub-leading term $c_q$.
The PEs are deeply tied to the concepts of inverse participation ratio (for ${q=2}$) and (multi)fractal dimension \cite{Stanley88}. They play a significant role in the Anderson~\cite{Evers00, Evers08, Rodriguez09, Rodriguez10}
and the many-body localization transitions~\cite{DeLuca13, Mace19, Luitz20, Pietracaprina21, Solorzano21, Monthus16, Serbyn17}. Moreover, the PEs distinguish phases of quantum matter~\cite{Stephan09, Stephan10,Alcaraz13, Stephan14, Luitz14, Atas12, Luitz14spectra, Luitz14improving, Lindinger19,Pausch21} without relying on the system specific observables.

In this work, we compute PEs across MIPTs in the stationary state of random stabilizer and Haar circuits. 
For stabilizer states, we show the absence of multifractal behavior, signaled by $S_q$ being $q-$independent, whereas for Haar circuits, the system state is multifractal for any finite measurement rate, with ${D_q<1}$ depending in a non-trivial fashion on $q$. 
In analogy to quantum phase transitions \cite{Zaletel11, Luitz14universal}, we find that the sub-leading term $c_q$ contains universal information about MIPTs. We provide an analytical understanding of those results by mapping the calculation of PEs onto a classical partition function~\cite{jian2020measurementinduced,bao2020theory,vasseur2019entanglement,zhou2019emergent,lopezpiqueres2020meanfield,fan2021selforganized} and show that our conclusions apply to a large class of quantum circuits.

\paragraph{Random quantum circuits. } We consider random quantum circuits with local projective measurements acting on $L$ qubits in the geometry shown in Fig.~\ref{fig1}  (periodic boundary conditions are assumed). The two-qubit gates $U_{j, j+1}$ are sampled uniformly from the Clifford group for stabilizer circuits and drawn randomly from the Haar distribution on $SU(4)$ for the Haar circuits. 
Each timestep consists of a layer of projective measurements onto the $z$ component of the spins, performed on each site ${j=1,\ldots,L}$ with probability $p$, and of a layer of unitary gates.
The system is in an error correcting phase for $p<p_c$ and in a quantum Zeno phase for $p>p_c$, with the critical rate identifying the MIPT ($p^S_c = 0.1593(5)$ for stabilizer circuits~\cite{li2019measurementdriven,gullans2020dynamical}, while $p^H_c = 0.17(1)$ for Haar circuits~\cite{zabalo2020critical}).

We initialize the circuit in the product state ${\ket{\psi_0}}$ ${=\ket{0,\ldots,0}}$ and calculate its time evolution for $t$ layers~\cite{footnote_init_state}. 
Stabilizer circuits are simulated in time polynomial in $L$ (up to $L\leq 880$) using the package Stim~\cite{Gidney21} and employing the ideas in~\cite{Gottesman98,aaronson2004improved, Koenig14}.
For Haar circuits, we perform exact simulations in the full Hilbert space, up to $L \leq 24$ qubits.
In both cases, we compute the PEs for each quantum trajectory $S_q(\ket{\psi_{\mathrm{ \textbf{m}}} })$ specified by the realization $\textbf{m}$, and consider the mean value. 
For the stabilizer circuits PEs, the average is over the times $2L\leq t\leq 22L$ and over $\mathcal{N} \ge 20000$ realizations, while, in the Haar case, we average over $2L \leq t \leq 1000 $, and over $\mathcal{N}\ge 4000\ (1000)$ circuit realizations for $L< 24 \ (L=24)$.

\paragraph{Participation entropy of stabilizer states. }
For a stabilizer state $\ket{\psi}$ on $L$ qubits there exists $L$ independent Pauli strings $g_j$ that \textit{stabilize} $\ket{\psi}$ \cite{nielsen2002quantum}: ${ g_j|\psi\rangle = |\psi\rangle }$. The  $g_j$ generate a group $\mathcal G$ and can be written as
${ g_j=e^{i\pi\phi_j} \prod_{k=1}^L X_k^{n_k^j}Z_k^{m_k^j} }$ where $X_k$, $Z_k$ are Pauli matrices acting on $k$-th qubit and $n_k^j,\ m_k^j,\ \phi^j$ are equal to $0$ or $1$. Hence, the stabilizer state $\ket{\psi}$ is uniquely determined by the matrices ${ M_X = [n_k^j] }$, ${ M_Z = [m_k^j] }$ and the vector of phases ${ \Phi=[\phi_j] }$.
Its density matrix reads $\rho = |\psi\rangle\langle \psi| = 2^{-L}\sum_{g\in \mathcal{G}} g$ and the PEs \eqref{sqdef} are 
\begin{equation}
\label{eq:SqST}
	{S}_q =  \frac{1}{1-q} \log_2 \left[ \mathrm{\sum}_{\vec{\sigma}}
	\left( 2^{-L}  \mathrm{ \sum }_{g \in \mathcal{G}} \langle \vec{\sigma}|  g |\vec{\sigma}\rangle \right) ^q \right].
\end{equation}
For specificity, we choose the eigenbasis of $Z_k$ operators (the $Z$ basis) as the basis $\{ \ket{\vec{\sigma}}\}$. Then, the matrix element $\langle \vec{\sigma}|g|\vec{\sigma}\rangle$ is non-zero only if the Pauli string $g$ contains no $X_i$ operators.
Such Pauli strings form a subgroup ${\mathcal{G}'\subset \mathcal{G}}$, generated by ${g'_i = \prod_{j=1}^L g^{a^j_i}_j}$, with $\vec{a}_i$ being the solutions of  ${M_X \vec{a}=0}$ over the field $\mathbb{Z}_2$.

The independence of $g_j$ implies that there are  $r=L-\mathrm{rk}_{\mathbb{Z}_2} M_X$ independent generators of $\mathcal{G}'$ (where $\mathrm{rk}_{\mathbb{Z}_2} M_X$ is the rank of the matrix $M_X$ over $\mathbb Z_2$), hence $|\mathcal{G}'|=2^r$. From~\eqref{eq:SqST}
we obtain that 
\begin{align}
	{S}_q = \frac{1}{1-q}\log_2 \left[\frac{2^{qr}}{2^{qL}}  \sum _{\vec{\sigma}}\left( \sum_{g'\in  \mathcal{G}'} \frac{\langle  \vec{ \sigma }|   g'| \vec{\sigma}\rangle}{2^r} \right)^q \right].\label{magic}
\end{align}
The operator $P_{\mathcal{G}'} \equiv 2^{-r} \sum_{g'\in \mathcal{G}'} g'$ is a projector onto a subspace stabilized by the elements of $\mathcal{G}'$. 
By definition, the elements of $\mathcal{G}'$ are diagonal in the the basis $\{|\vec{\sigma}\rangle\}$, hence 
$\bra{\vec \sigma}P_{\mathcal{G}'}  \ket{\vec \sigma} $ can be only $0$ or $1$,
allowing to change the order of sums in~\eqref{magic}. 
Since the Pauli matrices are traceless, $\mathrm{\sum} _{\vec{\sigma}} \langle \vec{\sigma}|   g' |\vec{\sigma}\rangle$ is non-vanishing and equal to $2^L$ only if $g'$ is the identity operator.
We conclude that  
\begin{align}
\label{eq:part4}
 S \equiv {S}_q = \mathrm{rk}_{\mathbb{Z}_2}(M_X).
\end{align}

Crucially, the result \eqref{eq:part4} holds for an arbitrary stabilizer state and is independent of $q$, implying that wave-functions of stabilizer states do not exhibit multifractality. This holds true for generic eigenbases of Pauli strings~\cite{suppl} and reminisces about the fact that for stabilizer states all R\'enyi entanglement entropies are equal~\cite{hamma2005bipartite, hamma2005ground}. The formula \eqref{eq:part4} has a simple interpretation. If $M_X=0$, the stabilizer state $\ket{\psi}$ is fully localized in the $Z$ basis and the PE is vanishing. 
Each linearly independent row of $M_X$ corresponds to a generator $g_j$ that delocalizes $\ket{\psi}$ in the $Z$ basis, incrementing PE by one.

\begin{figure}
	\centering
	\includegraphics[width=\linewidth]{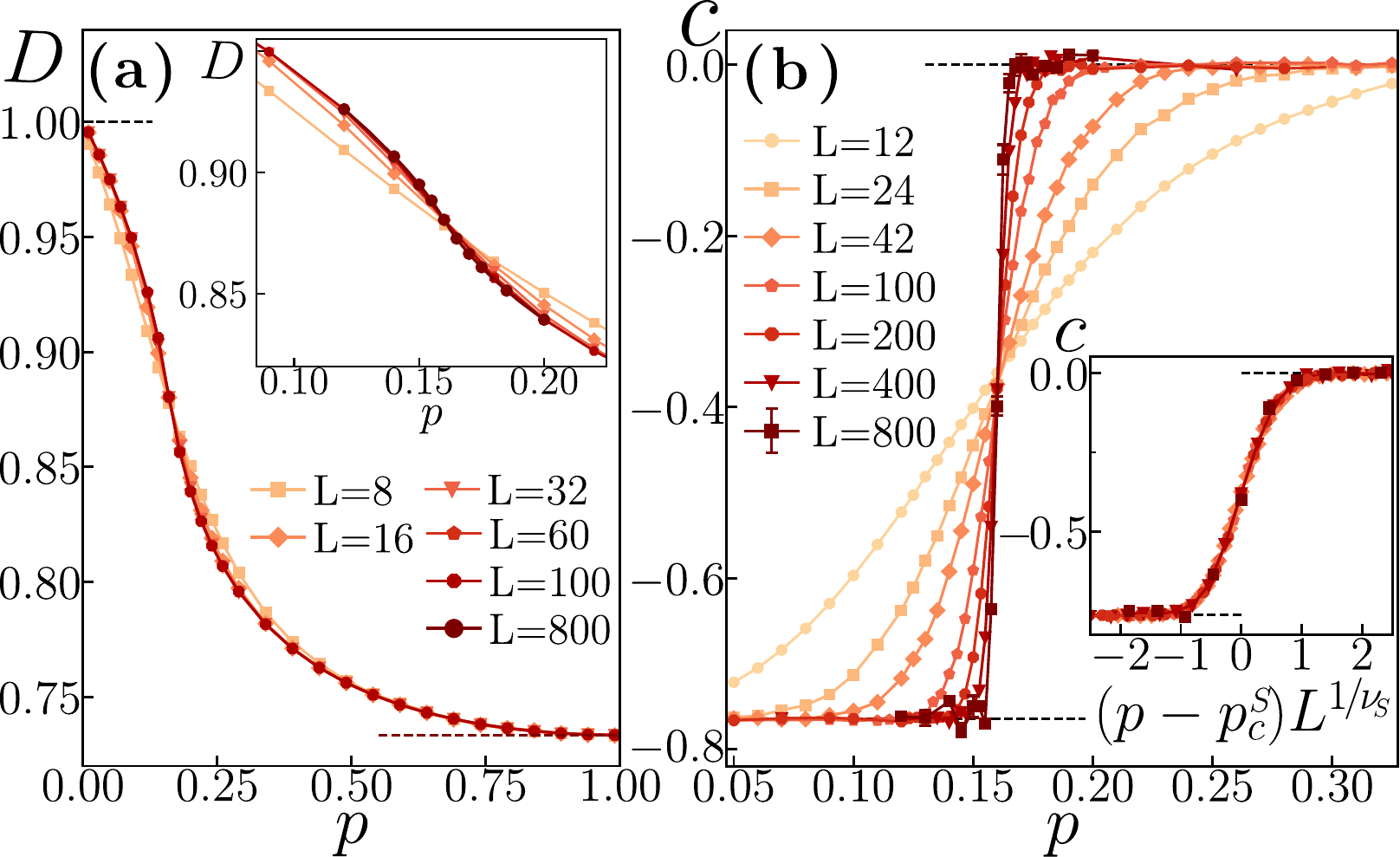} 
	\caption{\label{fig:2} 
	The fractal dimension $D$ \textbf{(a)} and the sub-leading term $c$ \textbf{(b)} at a measurement rate $p$ for system size $L$. Insets: \textbf{(a)} -- the vicinity of the critical point, \textbf{(b)} -- data collapse for $c$ (for $L\geq 42$) with $p^S_c=0.160(2)$ and $\nu_S=1.28(5)$. The dashed lines signify the $p \to0,1$ limits.
	}
\end{figure}

\paragraph{Stabilizer circuits.} We now turn to investigation of the MIPT in the Clifford circuits and calculate the PE ~\eqref{eq:part4} \cite{footnote_calc}. 
The fractal dimension $D$ and the sub-leading term $c$ are obtained from linear fits ${S(L_0)=D L_0 + c}$. To obtain $D(L)$ and $c(L)$ in a system size resolved manner, we use three chain lengths in the fit: $L_0=L-\delta L, L, L+\delta L $, where $\delta L = 2$ for $L<40$ and $\delta L = L/10$ for $L>40$.

The fractal dimension $D$ is shown in Fig.~\ref{fig:2}(a). The wave function is fully delocalized (${D\to 1}$) over the Hilbert space only for ${p\to 0}$. For ${p>0}$, we observe a fractal scaling of the wave function (${D<1}$), and that the fraction of Hilbert space occupied by the wave function decreases monotonously with the measurement probability $p$. 
The numerical results suggest that $D$ collapses on a limiting curve $D(\infty)$ with a flex point at $p=p_c$ for $L \to \infty$~\cite{suppl},
exhibiting similar behavior to that of the fractal dimension at equilibrium quantum phase transitions~\cite{Luitz14}.

Crucially, the universal information about the MIPT is completely encoded in the sub-leading term $c(L)$, which acts as an order parameter for the system.
Indeed, $c(L)$ approaches a step function, with discontinuity at the MIPT, for $p=p^S_c$ (Fig.~\ref{fig:2}~(b)).
Employing a scaling ansatz $c = f[  (p-p^S_c) L^{1/\nu_S} ]$ we find the data collapse on a universal curve with  $p^S_c=0.160(2)$ and $\nu_S=1.28(5)$ in agreement with earlier results based on entanglement measures~\cite{li2019measurementdriven}. 

The behavior of $D$ and $c$ in the ${p\to 0}$ and ${p\to 1}$ limits, highlighted with dashed lines in Fig.~\ref{fig:2},
can be easily understood.
In the limit of no measurements, we observe that our local Clifford circuit acts as a global $L$-qubit Clifford gate~\cite{Maslov18, Berg21} 
for which we analytically obtain $S=\sum_{n=1}^L \left( 1-(2^n+1)^{-1}\right) = L + c^\mathrm{stab}$, 
where $c^\mathrm{stab} \approx -0.7645$.
In the opposite limit of ${p\to 1}$, PE is vanishing after the layer of measurement. Subsequently, PE increases due to the layer of $L/2$ 2-qubit gates
yielding $S=D L$  with $D=11/15$. 
(See \cite{suppl} for details of these derivations, based on counting arguments and on properties of the Clifford group.)
Our numerical analysis shows that the sub-leading term $c$ matches these limiting values: $c\approx -0.7645$ for $p<p^S_c$ and $c = 0$ for $p>p^S_c$.

\begin{figure*}[ht]
	\centering
	\includegraphics[width=\linewidth]{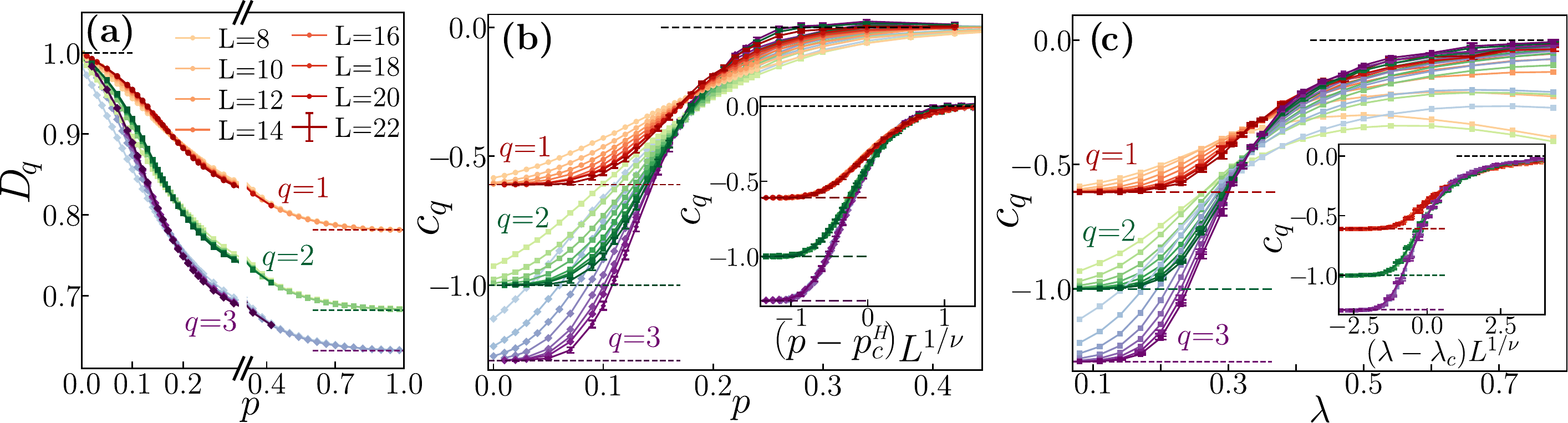} 
	\caption{\label{fig:3} 
The multifractal dimensions $D_q$ \textbf{(a)} and the sub-leading term $c_q$ \textbf{(b)} at  a measurement rate $p$ for the Haar circuit. Inset in \textbf{(b)}: collapse of $c_q$ at MIPT with $p^H_c=0.166(5)$ and $\nu=1.4(1)$. The $c_q$ for Floquet circuits \textbf{(c)} at a measurement strength $\lambda$. Inset in \textbf{(c)}: collapse for $\lambda_c=0.320(8)$, $\nu=1.3(1)$. In all the figures ${8\le L\le 22}$ and ${q=1,2,3}$.  
The dashed lines show the analytical predictions for $p\to0,1$ \textbf{(a-b)} and for $\lambda\to 0,1$ \textbf{(c)} limits.
	}
\end{figure*}

\paragraph{Haar circuits.} We investigate the MIPT by considering the PEs for $q=1,2,3$ and focusing on the $Z$ basis. 
To obtain $D_q$ and $c_q$, we consider the linear fits $S_q(L_0)=D_q L_0 + c_q$ for three system sizes $L_0=L-2, L, L+2$.
As captured by the non-trivial $q-$behavior of $D_q$ (see Fig.~\ref{fig:3}~(a)), the Haar circuits exhibit multifractality for any ${p>0}$. 
Furthermore, similarly to the stabilizer circuits, we observe a mild system size dependence of $D_q$ close to the MIPT, suggesting that $D_q(p)$ approaches a continuous curve in the  ${L\to \infty} $ limit with an inflection point at $p_c^H$.

The sub-leading term $c_q$ approaches a step function with increasing $L$ as shown in Fig.~\ref{fig:3}~(b). The value of $c_q$ is non-zero and $q$ dependent in the error correcting phase $p<p^H_c$, whereas it vanishes in the quantum Zeno phase $p>p^H_c$.
The sub-leading term $c_q$ collapses onto universal ($q$ dependent) curves 
upon rescaling  $ p \to (p-p^H_c) L^{1/\nu}$. 
The  parameters $p^H_c=0.166(5)$ and $\nu=1.4(1)$ are in agreement with the results in~\cite{zabalo2020critical} obtained from the scaling of the quantum mutual information. As for the stabilizer circuits, the sub-leading term $c_q$ acts as an order parameter for the MIPT. The values of $D_q$ and $c_q$ in the limiting cases $p\to 0$ ($p\to 1$) can be easily understood.
For ${p \to 0}$ and for ${t>2L}$, the PEs are well approximated by the PE obtained replacing the whole circuit by a matrix $U$ drawn with the Haar measure from the $SU(2^L)$ group.
This is expected since a sufficiently deep local Haar random circuit acting on $L$ qubits forms an approximate 
unitary $t$-design~\cite{Brandao16,Gross07, Harrow09, Brown10}.
Assuming that the initial state $\ket{\psi_{\mathrm{in}}}$ is the first basis state (which can be done due to translation-invariance of the Haar measure on $SU(2^L)$), we get that the PEs $S^{p=0}_q(L) =  \mathbb{E}_{U}  \log \left( \sum_{j=1}^{2^L} |U_{j,1}|^{2q} \right)  / (1-q)$, where $\mathbb{E}_{U}$ denotes the average with Haar measure.
The integral $I^q_N  \equiv \mathbb{E}_U \log_2 \left( \sum_{j=1}^N |U_{j,1}|^{2q}\right)$ for $U\in SU(N)$ can be easily evaluated numerically for small $N$~\cite{Puchala17}, and for ${N \gg 1}$ it is approximated by $I^q_N = (1-q) \log_2 N + \log_2(\Gamma(1+q) )$ 
~\cite{Weingarten78, Collins06,suppl}.
This yields $D_q=1$ and $c_q =(1-q)^{-1}\log_2( \Gamma(1+q) )$.
In the opposite limit, ${p\to 1}$, the layer of measurements leads to a collapse of the state of the system onto a product state $\ket{\psi_{\mathrm{ \textbf{m} } }} =\bigotimes_{i=1}^L \ket{e_i}$. A subsequent application of a layer of $L/2$ two-site gates $U_{i,i+1}$ leads to a state $\ket{\psi} = \bigotimes_{k=1}^{L/2} U_{2k-1,2k}\ket{e_{2k-1}e_{2k} }$
with $S^{p=1}_q = \frac{L}{2(1-q)} I^q_4$, so that in the ${p\to 1}$ limit, we get a non-zero $D_q$ and a vanishing sub-leading term $c_q$.

\paragraph{Replica approach to participation entropies. } In the following, we provide an intuitive interpretation of the extensive scaling of PEs, and of $c_q$ as an order parameter for the transition.
The entanglement entropy in Haar circuits can be calculated with a replica approach by means of a mapping to a classical statistical model~\cite{jian2020measurementinduced,bao2020theory,vasseur2019entanglement,zhou2019emergent,lopezpiqueres2020meanfield,fan2021selforganized}.
We generalize the Haar circuit to qudits with on-site Hilbert space dimension $d$ and
express the PEs \eqref{sqdef} in terms of partition functions of a classical 2D spin model, translating the MIPT to an equilibrium ordering transition~\cite{suppl}.
The bulk of the spin model is fixed by the spatiotemporal structure of the circuit (coinciding with the one implemented for the entanglement entropy computation \cite{jian2020measurementinduced}), and the PEs are fully encoded in the choice of boundary conditions. 
An immediate consequence is that both PEs and entanglement entropy encode the same critical properties of MIPTs. 
Furthermore, this mapping provides a direct link between the quantum phase transitions in spin models at equilibrium~\cite{Zaletel11, Luitz14universal}, and our non-equilibrium setup: in both cases $c_q$ is an order parameter for the transition.

The calculation of the PEs simplifies in the limit of large on-site Hilbert space dimension $d$. At leading order in $d$ and in the limiting cases of small and large measurement rate $p$, we obtain closed expressions:
the fractal dimension is given by
$D_q =\log_2 d - \frac{p^2}{2(1-q)} \log_2 \Gamma(1+q)$,
and the sub-leading term is $c_q = (1-q)^{-1}\log_2( \Gamma(1+q) )$ for $p\rightarrow0$ up to terms $O(p^4)$; 
$D_q=\log_2 d + \frac{1}{2} \frac{\log_2 \Gamma(1+q)}{1-q} + \frac{ (1-p)^2}{2(q-1)}\log_2 \Gamma(1+q)$,
$c_q=0$ for $p\rightarrow1$ up to terms $O((1-p)^3)$.
These results are perfectly in line with our numerical findings, indicating the presence of multifractality and showing that $D_q$ is a decreasing function of $p$, 
whereas $c_q$ is constant throughout the error-correcting and quantum Zeno phase. 
The replica calculation of the PEs provides also an intuition for the order-parameter behavior of $c_q$. For $p\to0$, we find that the non-zero value of $c_q$ originates from the correlation between spins that point in the same direction on all lattice sites. In contrast, for $p\to1$ the spins on different sites are fully uncorrelated, which results in a vanishing $c_q$. Heuristically, this observation (valid for the limiting cases) can be extended to the whole phase provided the thermodynamic limit is taken.  
The region $p<p_c$ corresponds to a ``ferromagnetic'' phase of the spin model~\cite{bao2020theory}, in which the correlation between spins throughout the $2D$ lattice leads to a non-zero value of $c_q$. Instead, $p>p_c$ corresponds to a disordered phase of the spin model, in which the spins are uncorrelated beyond a certain length scale, giving rise to the PE strictly proportional to system size $L$ and hence to the vanishing $c_q$.

 \paragraph{Universality.} To show the generality of our results, we analyze PEs across MIPT in various settings: i) a Clifford circuit with rank-2 measurements onto local Bell pairs,
ii) Haar circuit with on-site Hilbert space dimension $d=3$, iii) a Floquet circuit with alternating application of $U_F = e^{-i \sum_j X_j} e^{ -i \sum_j ( Z_j Z_{j+1}  + Z_j ) } $
and measurements $M_{j,\pm} = (1\pm \lambda Z_j)/(2+2\lambda^2)$ ($j \in [1,L]$) with probability $p=1$ and strength $\lambda$. In \cite{suppl} we detail our numerical results which are summarized as follows. The sub-leading term, $c_q$, plays the role of order parameter for MIPT in each of the cases i)-iii). For the Clifford circuit with rank-2 measurement, we found a collapse of $c_q$ with $p^{S_2}_c=0.692(2)$ (in agreement with \cite{li2018quantum}) and $\nu^{S_2}=1.30(3)$ consistent with $\nu^S$ according to the expectation that a local change of the measurement protocol does not alter universality class of MIPT.
For ii): Haar circuit with $d=3$, $c_q$ approaches a step function. The values of $c_q$ in the error-correcting phase are given by $c_q = (1-q)^{-1}\log_2( \Gamma(1+q) )$ match the results for $d=2$, $d\to \infty$ and, importantly, coincide with random matrix theory prediction \cite{Backer19} for systems with broken time reversal symmetry. 
The Floquet circuit iii) is composed of $U_F$ generated by local Hamiltonians without any randomness and of weak measurements $M_{j,\pm}$ with strength $\lambda$ that allows to tune the system across MIPT \cite{li2019measurementdriven}. Despite the differences with random quantum circuits, the behavior of $D_q$ and $c_q$ is analogous. The multifractal dimension $0<D_q<1$ depends in a non-trivial fashion on $q$ for $\lambda>0$ confirming that multifractality of wave-function can be generally expected in many-body systems with measurements. The sub-leading term $c_q$, shown in Fig.~\ref{fig:3}~(c), is an order parameter for MIPT and rescaling $\lambda \to (\lambda-\lambda_c)L^{1/\nu}$ leads to a collapse of data with $\lambda_c=0.320(8)$ and $\nu=1.3(1)$ consistent with entanglement measures \cite{suppl}.

\paragraph{Conclusions. } We have shown that the PEs of quantum circuits with local measurements allow for a full characterization of MIPTs, alternative to entanglement measures. 
The expression for the PE of stabilizer states exhibits absence of multifractality. This emphasizes  their simplicity as compared to generic many-body wave functions, and provides yet another perspective on the classical tractability of stabilizer circuits and Gottesman–Knill theorem.
The multifractal dimensions, the leading terms in system size scaling of the PEs, indicate that the generic wave-functions are multifractal in the presence of projective measurements in the system. 
The sub-leading term $c_q$ encodes universal information about MIPTs, acting as an order parameter
with value determined by whether the circuit is generic (e.g. Haar and Floquet circuits) or specifically fine-tuned (stabilizer circuits) and by gross features of quantum dynamics such as the dynamical phase or the presence of time reversal symmetry.
This raises general questions about the connection of $c_q$ to ergodicity of quantum many-body systems and, in particular, about behavior of $c_q$ across many-body localization transition (cf. \cite{Mace19}).
Our results highlight the importance of
the structure of many-body wave-functions for MIPTs, and can be used as a reference point in the vividly pursued investigations of MIPTs in many-body systems such as the Bose-Hubbard model \cite{tang2020measurementinduced, goto2020measurementinduced}, spin systems \cite{Dhar16,  lang2020entanglement, biella2021manybody, turkeshi2021measurementinduced_2, Botzung21, sierant2021dissipative} or free fermions \cite{cao2019entanglement, muller2021measurementinduced, alberton2021entanglement, minato2021fate, Zhang21}. 
Recent progress in stochastic sampling of wave functions in atomic as well as in solid-state platforms~\cite{Brydges19,Leseleuc19,Chiaro20,Pagano2020quantum,Scholl20,Ebadi20,Zeiher17,Veit21} shows that direct probing of PEs constitutes an experimentally viable alternative to entanglement measures \cite{gullans2020scalable}.
Random circuit sampling in superconducting quantum processors \cite{Arute19, Wu21} offers another possibility to study PEs experimentally (see \cite{suppl} for analysis of an experimental setup).
The multifractality at MIPTs was recently investigated also in a different context of: i) correlation functions \cite{zabalo2021operator} that exhibit multifractal scalings at MIPT in generic models 
ii) many-body wave-functions transformed to graph states  \cite{Iaconis2021multifractality} which can be mapped to Anderson models.

\begin{acknowledgments}
\textit{Acknowledgments.} We acknowledge collaboration on related subjects as well as enlightening conversations with M. Dalmonte, R. Fazio, M. Schir\'o, A. Biella,  G. Chiriacò, F. M. Surace, S. Sharma, G. Pagano, A. Scardicchio, and J. Zakrzewski. XT thanks S. Pappalardi for insightful discussions on related topics. PS acknowledges the support of  Foundation  for Polish   Science   (FNP)   through   scholarship START and support by PL-Grid Infrastructure. XT is partly supported by the ERC under grant number 758329 (AGEnTh), by the MIUR Programme FARE (MEPH), and by the European Union’s Horizon 2020 research and innovation programme under grant agreement No 817482 (Pasquans).
\end{acknowledgments}

\bibliographystyle{apsrev4-1} 
\normalem
%

\widetext
\clearpage
\begin{center}
\textbf{\large \centering Supplemental Material:\\ Universal behavior beyond multifractality of wave-functions at measurement--induced phase transitions}
\end{center}

\setcounter{equation}{0}
\setcounter{figure}{0}
\setcounter{table}{0}
\setcounter{page}{1}
\renewcommand{\theequation}{S\arabic{equation}}
\setcounter{figure}{0}
\renewcommand{\thefigure}{S\arabic{figure}}
\renewcommand{\thepage}{S\arabic{page}}
\renewcommand{\thesection}{S\arabic{section}}
\renewcommand{\thetable}{S\arabic{table}}
\makeatletter

\renewcommand{\thesection}{\arabic{section}}
\renewcommand{\thesubsection}{\thesection.\arabic{subsection}}
\renewcommand{\thesubsubsection}{\thesubsection.\arabic{subsubsection}}

\vspace{1cm}
The Supplemental Material contains:
\begin{enumerate}
    \item A discussion of the basis dependence of participation entropy for Haar and stabilizer circuits. This includes also a discussion on measuring the PE after a measurement layer, oppositely to the Main Text, where we discuss the PE computed before the measurement layer.
    \item The behavior of PEs in stabilizer and Haar circuits in the limits $p \to 0$ and $p \to 1$.
    \item A discussion on self-averaging, and the comparison with the multifractal analysis obtained through inverse participation ratio.
    \item The crossover of the fractal dimension in stabilizer circuits toward the limiting thermodynamic form.
    \item The replica calculation of the participation entropies, which in turns includes:
    \begin{itemize}
        \item details of the mapping of the circuit to a classical spin model;
        \item results for the limit of large on-site Hilbert space dimension $d$;
        \item cluster expansion results at $p\to0$ and $p\to1$.
    \end{itemize}
    \item Additional numerical analysis, on:
    \begin{itemize}
        \item Clifford circuits with rank-2 measurements;
        \item Haar circuits for $d=3$ qudits;
        \item Floquet quantum circuits with generalized measurements.
    \end{itemize}
    \item An experimental setup proposal on state-of-the-art and near term quantum computers.
\end{enumerate}

\section{S1 Basis dependence of the participation entropies}
\label{sec:bas}

As we noted in the Main Text, the participation entropies
\begin{align}
	S_q = \frac{1}{1-q} \log_2 \sum_{{\vec{\sigma}}} |\langle {\vec{\sigma}} |\psi \rangle|^{2q},
	\label{sqdefSUP}
\end{align}
are dependent on the choice of the many-body basis $\{ \ket{ \vec{\sigma}} \}$. Here, we describe this dependence for Haar circuits and stabilizer circuits, emphasizing the universal information about MIPT contained in the sub-leading terms $c_q$ in system size scaling of participation entropy $S_q=D_q L + c_q$.

\subsection{Haar random circuits}

The numerical results presented in the Main Text for the Haar circuits were obtained in the $Z$ basis (the eigenbasis of $Z_j$ operators) with participation entropies calculated after the layer of unitary gates.
However, within this prescription, the the results remain exactly the same in the generic eigenbasis of local operators $U^*_j Z_j U_j$ where $U_j$ are arbitrary (fixed at a given site) unitary matrices from $SU(2)$ group. This is an immediate consequence of the translation-invariance of the Haar measure on the $SU(4)$ group from which the $2$ qubit gates $U_{i,i+1}$ are drawn. 

\begin{figure}[h!]
	\centering
	\includegraphics[width=\linewidth]{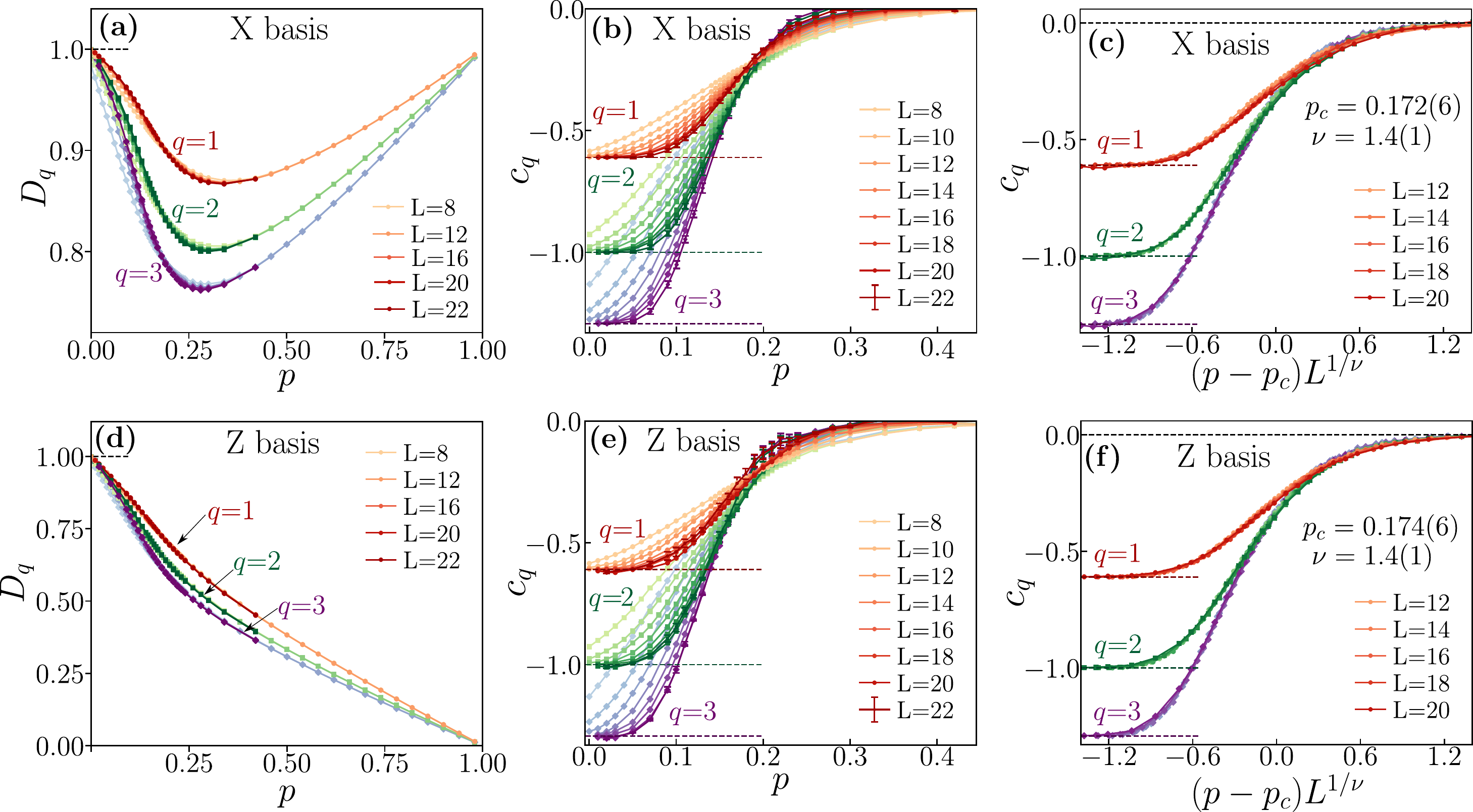}
	\caption{\label{fig:S1} Multifractality at MIPT in the Haar circuit: basis dependence. Participation entropies calculated after the measurement layer. \textbf{(a), (b)}: the fractal dimension $D_q$ and the sub-leading term $c_q$ as function of the measurement probability $p$ calculated in the $X$ basis (the eigenbasis of $X_k$ operators), \textbf{(c)}: collapse of the the sub-leading term upon rescaling ${p \to (p-p_c) L^{1/\nu}}$. \textbf{(d), (e), (f)}: Results of analogous computations performed in the $Z$ basis (eigenbasis of the $Z_k$ operators).}
\end{figure}

If instead the participation entropies are calculated after the layer of measurements, the freedom of local rotations is no longer present since the basis associated with the projectors $P_{\pm}=(1\pm Z)/2$ becomes distinct. The results for that case are shown in Fig.~\ref{fig:S1}.

The fractal dimensions $D_q$, shown in Fig.~\ref{fig:S1}\textbf{(a)},~\textbf{(d)} coincide only in the limit $p\to0$, when the wave-function is fully delocalized ($D_q \to 1$). Close to the critical measurement probability, we see a crossing point of $D_q$ curves for various $L$. 
In the $Z$ basis, with the increase of $p$, the wave function is extended over decreasing fraction of Hilbert space and gets localized ($D_q=0$) in the $p\to1$ limit. Localization of the wave-function in the $Z$ basis is equivalent to its full delocalization in $X$ basis, as indicated by the value of $D_q$ for $p\to1$ in $X$ basis. The results for $D_q$ show that the behavior of fractal dimensions is strongly dependent on the details of the protocol of their calculation. However, in all of the cases considered, $D_q$ demonstrate the multifractal scalings of wave-functions in presence of local projective measurements. 

The sub-leading terms $c_q$, shown in Fig.~\ref{fig:S1}\textbf{(b-e)} are independent of the choice of basis, containing the universal information about MIPT: the finite-size collapses with the scaling form $c_q = f[  (p-p_c) L^{1/\nu} ]$ yield $p_c$ and $\nu$ that agree, within estimated error bars, regardless of the choice of basis.

\subsection{Stabilizer circuits}

\begin{figure}[h!]
	\centering 
	\includegraphics[width=0.7\columnwidth]{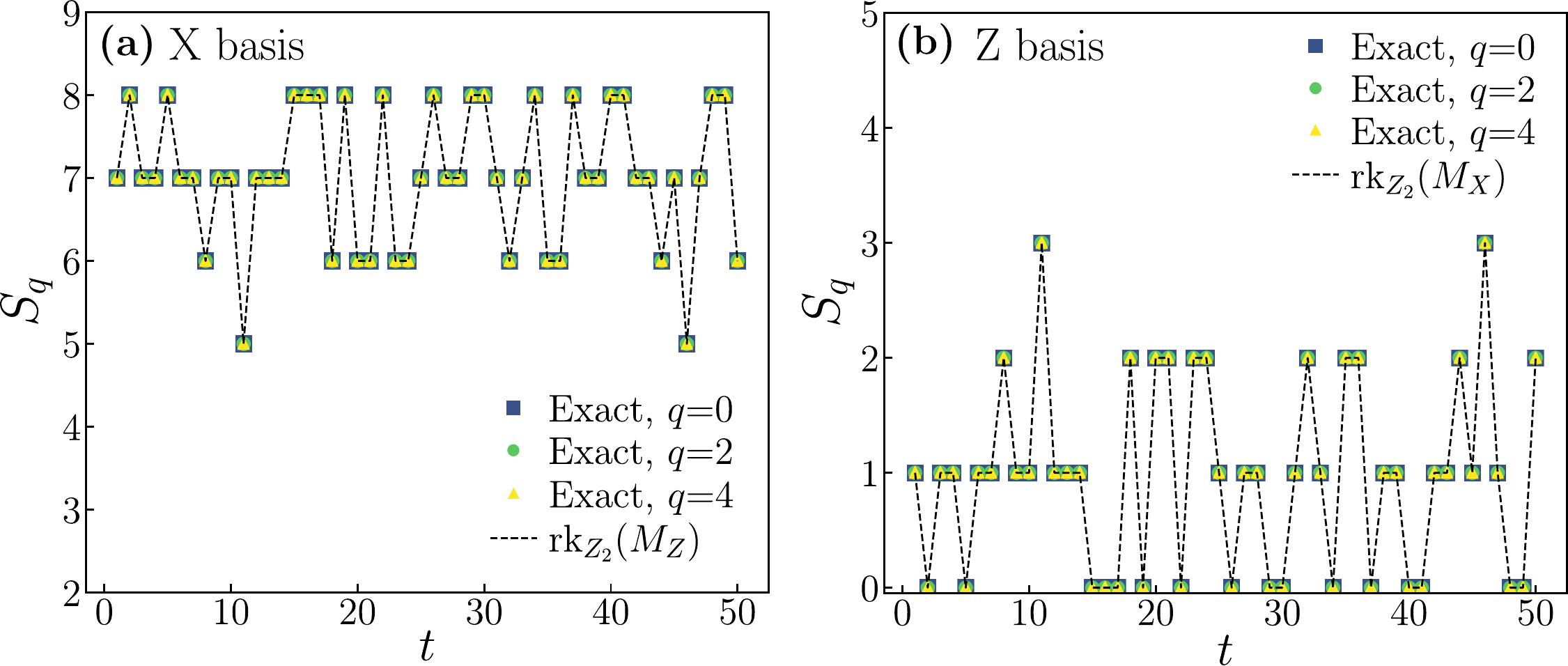}
	\caption{\label{fig:s2} \emph{Ab initio} test of the expressions \eqref{supSQ1}, \eqref{supSQ2} for participation entropy of stabilizer states against participation entropy obtained from the coefficients of wave-function in respective many-body basis. Results for the Clifford circuit of $L=8$ qubits at $p=0.3$, with the participation entropy computed after the measurement layer. \textbf{(a)}: results for X basis; \textbf{(b)} results for Z basis.}
\end{figure}

The expression 
\begin{align}
 {S}_q = \mathrm{rk}_{\mathbb{Z}_2}(M_X).
 \label{supSQ1}
\end{align}
was derived in the Main Text for participation entropy for the $Z$ basis. An analogous formula can be derived for the X basis
\begin{align}
 {S}_q = \mathrm{rk}_{\mathbb{Z}_2}(M_Z).
 \label{supSQ2}
\end{align}
(This results can be extended to the eigenbasis $\{\sigma\}$ of general Pauli strings, where the $M_\mathrm{\{\sigma\}}$ is extracted through the solution of a linear set of equations in $\mathbb{Z}_2$. See Main Text). 
Fig.~\ref{fig:s2} shows the results of a direct test of the formulas \eqref{supSQ1}, \eqref{supSQ2}.
Both results show that $S_q$ is independent of the parameter $q$ and hence the wave functions of stabilizer states are not multifractal.  The equations \eqref{supSQ1}, \eqref{supSQ2} hold in the respective eigenbases of operators $X_k$ and $Z_k$ that determine the Pauli strings $g_j$ specifying the stabilizer state. 
In general, basis operators related by a Clifford unitary transformation would preserve the absence of multifractality for the wave function of stabilizer states. Instead, generic unitary transformation $U^*_j$ (\textit{e.g.} Haar distributed) would induce a non-trivial $q$ dependence and multifractality.

\section{S2 Participation entropy in stabilizer and Haar circuits in the limits of high/low measurement rate}
\label{sec:int}

In this section we provide details of derivations of formulas for PEs in both stabilizer and Haar circuits in the limiting cases of vanishing measurement probability $p \to 0$ and large measurement probability $p \to 1$.

Our derivation in the limit of no measurements ($p \to 0$), is based on the observation that a local stabilizer (Haar) circuits acts as global $L$-qubit Clifford (Haar) gate. This can be expected for stabilizer circuits since an arbitrary Clifford operation on $L$ qubits can be expressed in terms of a linear in $L$ number of 2-qubit gates in nearest-neighbor architecture~\cite{Maslov18, Berg21}. For Haar circuits, such a behavior is also expected since a sufficiently deep local Haar random circuit acting on $L$ qubits forms an approximate unitary $t$-design~\cite{Brandao16}. In contrast, for the large measurement probability $p \to 1$ our derivations are based on the fact that for both stabilizer and Haar circuits the state is collapsed onto a Fock state by the layer of $L$ measurements of the $Z_i$ operators.

To progress in the case of stabilizer circuits, we note that any operator $U_C$ from the Clifford group can be written as $U_C = F_1HF_2$ \cite{Bravyi21} where $H$ is a layer of Hadamard gates, and where $F_1$, $F_2$ reshuffle vectors of the $Z$ basis up to a phase. Hence, $H$ is the only term that influences the value of PE: For an initial product state, each Hadamard gate produces an equal superposition of $\ket{0}$ and $\ket{1}$ states, increasing the value of PE by $1$. Therefeore, the average PE after the action of $U_C$ is equal to the expected number of Hadamard gates in the Clifford operation. If $U_C$ is an $L$-qubit Clifford gate, the probability of having a Hadamard gate acting on qubit $n$ is $p_n=2^n/(2^n+1)$ \cite{Bravyi21}. The PE is given by  $S=\sum_{n=1}^L \left( 1-(2^n+1)^{-1}\right) = L + c^\mathrm{stab}$, 
where the sum $c^\mathrm{stab}=-\mathrm{\sum}_{n=1}^L (2^n+1)^{-1}$ converges rapidly with $L$ to $-\sum_{n=1}^{\infty} (2^n+1)^{-1} = -1 + \psi_{1/2}(1-i \pi/\ln(2))/\ln(2)\approx -0.7645$. (Here $\psi_q$ is the $q$-th polygamma function).
In the opposite limit of ${p\to 1}$, the state is collapsed on a single Fock state and the PE is vanishing after the layer of measurements. Subsequently, PE increases due to the layer of $L/2$ 2-qubit gates. Each of them has the average number of Hadamard gates equal to $22/15$ (as implied by the expression $p_n=2^n/(2^n+1)$), hence PE is 
given as $S=D L$  with $D=11/15$ and a vanishing sub-leading term $c=0$. 
The numerical analysis from the Main Text shows that the sub-leading term $c$ in stabilizer circuits matches the limiting values for $p \to 0$ and $p \to 1$ throughout the entire quantum error correcting and quantum Zeno phases.

For the Haar circuits, in the $p \to 0$ limit, we calculate the PEs by replacing the whole circuit by a matrix $U$ drawn with the Haar measure from the $SU(2^L)$ group. Using the translational invariance of Haar measure on $SU(2^L)$, the PEs are given by
$S^{p=0}_q(L) =  \mathbb{E}_{U}  \log \left( \sum_{j=1}^{2^L} |U_{j,1}|^{2q} \right)  / (1-q)$.
The integral 
\begin{align}
   I^q_N  \equiv \mathbb{E}_U \log_2 \left( \sum_{j=1}^N |U_{j,1}|^{2q}\right)
\end{align} where $U\in SU(N)$ and $\mathbb{E}_{U}$ denotes the average with Haar measure over the unitary group  $SU(N)$ can be easily evaluated numerically by noting that $|U_{j,1}|^{2}$ are uniformly distributed on a standard $N$--simplex $\Delta^{N} = \{ \lambda \in \mathbb{R}^{N}: \lambda_i \leq 0, \sum_{i=1}^{N}\lambda_i = 1\}$ \cite{Puchala17}. This can be done with Monte Carlo method by noting that Dirichlet distribution $\mathrm{Dir}(\lambda_1,\ldots,\lambda_N;a_1=1,\ldots, a_N=1)$ is a uniform distribution over $\Delta^{N}$.
The calculation simplifies for $N \gg 1$, when the real and imaginary parts $U_{j,1}$ behave as uncorrelated Gaussian random variables with a vanishing mean and variance equal to $N^{-1}$ \cite{Weingarten78, Collins06}.  Consequently, $t_j=|U_{j,1}|^2$ are independent random variables distributed with the exponential distribution $P(t_j)=N e^{-N t_j}$. Introducing a variable $w = \sum_{j} t_j^q$ we note that $w$, as a sum of $N \ll 1$ independent random variables, is normally distributed with mean proportional to the Gamma function $\mu = N^{1-q} \Gamma(1+q)$ and standard deviation $\sigma = N^{1-2q} \sqrt{ \Gamma(2q+1)-\Gamma(q+1)^2 } $. For $N \gg 1$, we have $\sigma \ll \mu$, and  $\log_2(w) \approx \log_2(\mu)$, which yields the results 
\begin{align}
I^q_N = (1-q) \log_2 N + \log_2(\Gamma(1+q) ) + O\left(\frac{1}{N}\right).
\end{align}
This leads to the following expression for the PEs:
\begin{align}
S^{p=0}_q(L) = L + (1-q)^{-1} \log_2 \Gamma(1+q), 
\label{eq:sqUNI2}
\end{align}
which approximates well the results already for moderate size circuits (e.g. $L=8$, since $2^L \gg 1$).
From \eqref{eq:sqUNI2} we get that the fractal dimension $D_q=1$ and the sub-leading term $c_q= (1-q)^{-1} \log_2 \Gamma(1+q)$. In contrast, as we argue in the Main Text, the PEs in the limit $p \to 1$ are given by $S^{p=1}_q = \frac{L}{2(1-q)} I^q_4$ and the sub-leading term is vanishing. The expressions derived here for the sub-leading term $c_q$ in the $p\to0$  ($p\to1$) limit match the numerical results in the entire quantum error correcting (quantum Zeno) phase.

\section{S3 Inverse Participation Ratio and self-averaging}
Another quantity of employed in the literature to probe multifractal properties of the wave function, particularly relevant for Anderson transitions~\cite{Evers08}, is the Inverse Participation Ratio (IPR), defined as
\begin{equation}
    \mathrm{IPR}_q \equiv 2^{(1-q) S_q}=\sum_{\vec{\sigma}} |\langle \vec{\sigma}|\psi\rangle|^{2q}.\label{eq:defipr}
\end{equation}
The typical value of IPR, $\mathrm{IPR}^\mathrm{typ}_q \equiv \exp{ \left \langle \log \mathrm{IPR}_q \right \rangle}$ (where $\left \langle . \right \rangle$ denotes the average over ensemble of states $\ket{\psi_{\mathrm{\textbf{m}}}}$) is functionally dependent on the average value of PEs: $\frac{1}{1-q} \log( \mathrm{IPR}^\mathrm{typ}_q ) = \left \langle S_q \right \rangle$. Therefore, the analysis of the average PEs, conducted in the Main Text, is equivalent to an analysis of the typical value of the IPR. In this section we investigate behavior of the \textit{average} IPR: $\,\left \langle \mathrm{IPR}_q \right \rangle$. The average IPR can be written as~\cite{Evers08}: $
\left \langle \mathrm{IPR}_q \right \rangle = a_0 2^{(1-q)\tilde{D}_q L}$
where $\tilde{D}_q$ is a fractal dimension, $a_0$ is a proportionality factor, we have neglected sub-leading terms and taken into account that the dimension of the Hilbert space is $2^L$. Taking a logarithm of this expression we observe that 
\begin{equation}
(1-q)^{-1} \log_2{\left \langle \mathrm{IPR}_q \right \rangle} = \tilde{D}_q L+(1-q)^{-1} \log_2 a_0
\equiv \tilde{D}_q L+ \tilde{c}_q,
\label{eq:ipr3}
\end{equation}
where we have introduced $\tilde{c}_q=(1-q)^{-1} \log_2 a_0$. We analyze this expression for stabilizer circuits in system size resolved manner, similarly as for the average PEs in the Main Text.

\begin{figure}[h]
	\centering
	\includegraphics[width=0.6\linewidth]{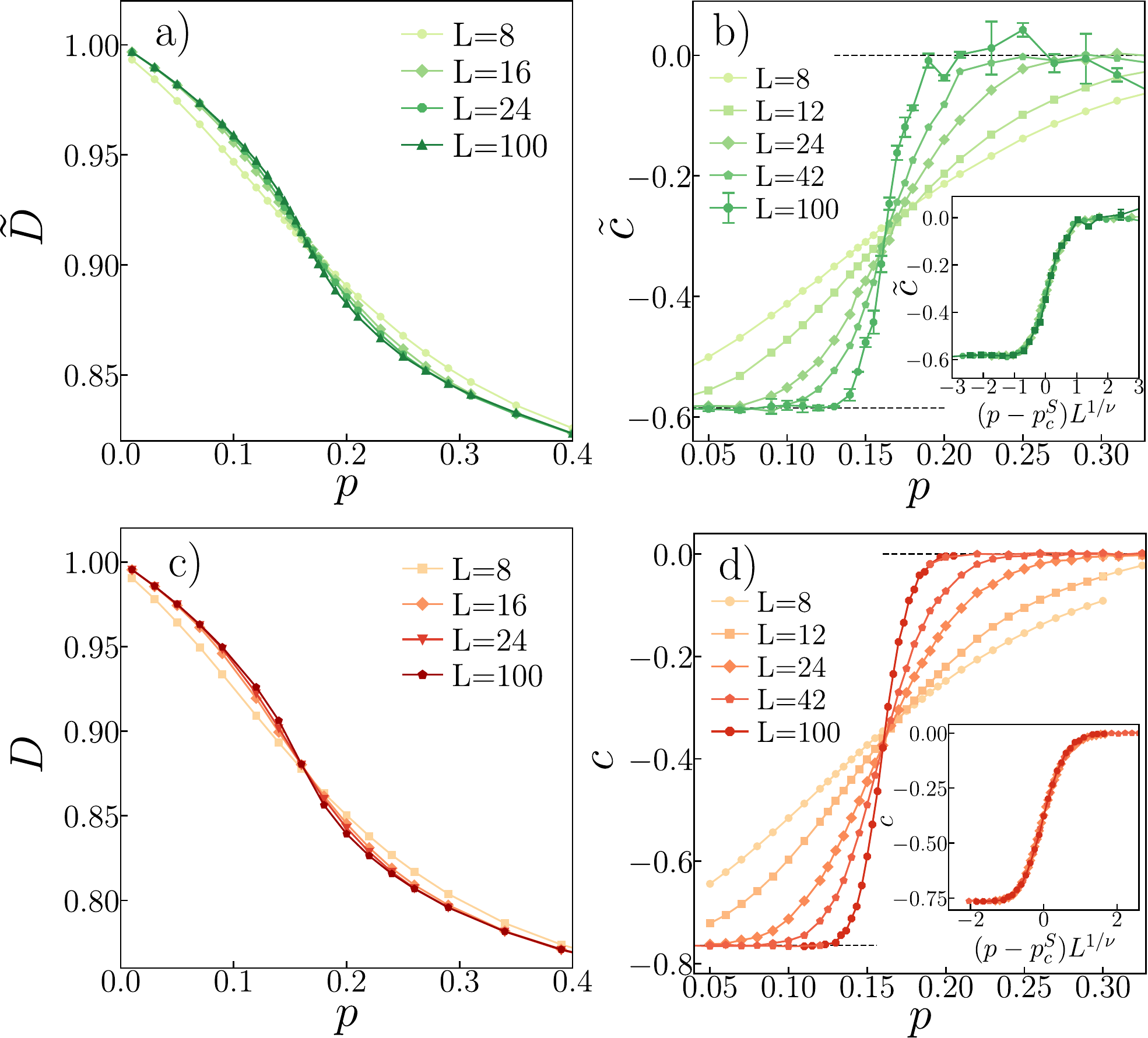}
	\caption{\label{fig:rIPR} Comparison of $\tilde{D}$ (a) and $\tilde{c}$ (b) obtained from the system size dependence of IPR: $(1-q)^{-1} \log_2{\left \langle \mathrm{IPR}_q \right \rangle} = \tilde{D}_q L+(1-q)^{-1} \log_2 a_0$ with  $D_q$ (c) and $c_q$ (d) obtained from the PEs: $S = D L + c$. The information about MIPT encoded both in the IPR and in the PE is the same (although there are minor differences: for instance, the values of $\tilde c$ and $c$ are different in the error-correcting phase.}
\end{figure}

For stabilizer states, 
using the result that  $S \equiv {S}_q = \mathrm{rk}_{\mathbb{Z}_2}(M_X)$ and \eqref{eq:defipr}, we immediately find that $\mathrm{IPR}_q = 2^{(1-q)\mathrm{rk}_{\mathbb{Z}_2}(M_X)}$ in the $Z$ basis. 
We report the numerical results for Clifford circuits in Fig.~\ref{fig:rIPR}, where we compare the fractal dimension $\tilde{D}$ ($D$) and the sub-leading term $\tilde{c}$ ($c$) extracted from the IPR (PE). The results are averaged over more than $20000$ circuit realizations.
Despite mild deviation in the precise value of $D$ and $c$ (compared to $\tilde{D}$ and $\tilde{c}$) the qualitative features and the critical scaling at the MIPTs are the same. This suggests that the system exhibits a qualitative self-averaging and the information extracted from the average IPR and typical IPR (equivalent to PEs) is the same.
Both $D_q$ and $\tilde{D}_q$ exhibit similar, mild finite size effects at the MIPT as visible in  Fig.~\ref{fig:rIPR}~a),~c). Moreover, the sub-leading term $\tilde{c}_q$ approaches a step function at MIPT with increasing $L$, analogously to $c_q$, as shown in Fig.~\ref{fig:rIPR}~b),~d). For both cases $p_c=p_c^S=0.160(5)$ and $\nu=1.30(9)$ (note that the error bars are larger than in the Main Text do tue a smaller interval of system sizes considered). 
Similar considerations extend also to Haar circuits, but we do not report the numerical results for the sake of readibility. 
We conclude by noting that the exponential in Eq.~\eqref{eq:defipr} gives rise to larger errors in the data, which is particularly visible in the sub-leading term $\tilde{c}_q$ shown in Fig.~\ref{fig:rIPR}~b). This justifies our choice of considering the participation entropy instead of the IPR to investigate the multifractal and the MIPTs in the quantum circuits of interest.

\section{S4 Crossover behavior for the fractal dimension of stabilizer circuits}
\begin{figure}
    \centering
    \includegraphics[width=0.35\columnwidth]{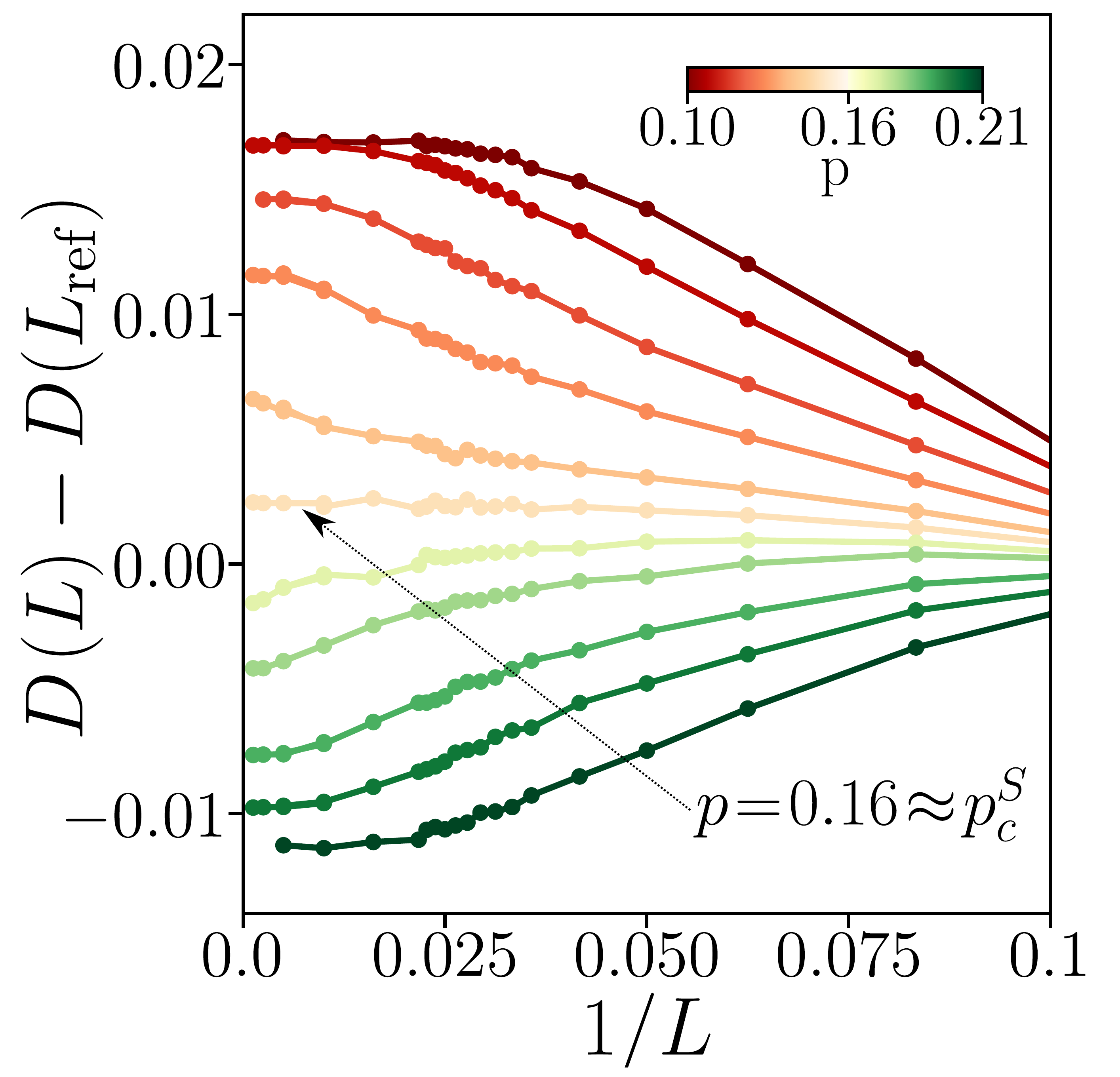}
    \caption{Crossover behavior for the stabilizer circuit fractal dimension. We fix $L_\mathrm{ref}=8$ and notice that at a fixed rate $p$ the value of $D(L)$ saturates for $L>L^*$.}
    \label{fig:crossoer}
\end{figure}

The numerical results suggest that the fractal dimension $D$ collapses on a limiting curve $D(\infty)$ with a flex point at $p=p_c$ for $L \to \infty$. 
This is presented in Fig.~\ref{fig:crossoer}, where we analyze the $L$ dependence of $D$ considering $D(L)-D(L_{\mathrm{ref}})$ and taking the reference system size $L_{\mathrm{ref}}=8$.
Beyond a length scale $L^*$, the value of $D(L)$ is a good estimate for $D(\infty)$ (e.g. $L^*\approx40$ at $p=0.1$). 
The value of $L^*$ increases as $p$ gets closer to $p_c$; however, outside of the interval of $p\in(0.15, 0.17)$, the signs of saturation of $D(L)$ are clearly visible.
This suggests that the system size dependence in $D(L)$ are crossover effects which are washed away in the thermodynamic limit and that the $D(\infty)$ possesses a flex point at $p=p^S_c$, 
exhibiting similar behavior to that of the fractal dimension at equilibrium quantum phase transitions~\cite{Luitz14}. However, our numerical results do not allow us to unambiguously exclude the possibility of a non-trivial asymptotic scaling of $D(L)$.

\section{S5 Mapping to a classical statistical mechanics model}
\label{sec:rep}

In this section, we detail the computation of the average participation entropies (PEs) for a Haar random circuit acting on $d$-dimensional qudits. We first introduce the mapping from the random circuit to a classical statistical model and then obtain closed expressions for the PEs in the limit of $d\gg 1$. 

\subsection{From random circuits to classical lattice spin models}
\label{sec:rep1}

This subsection follows~\cite{jian2020measurementinduced,bao2020theory,vasseur2019entanglement,zhou2019emergent,lopezpiqueres2020meanfield,fan2021selforganized}, where the mapping from the Haar hybrid circuit to a classical statistical mechanics model has been already considered. 
Each quantum trajectory $|\psi_\mathbf{m}\rangle = {K_\mathbf{m}|\psi\rangle}/||K_\mathbf{m}|\psi\rangle||$ is specified by the action of the non-unitary operator $K_\mathbf{m}$ describing the circuit realization $\mathbf{m}$ on an initial state $|\psi\rangle$.
We are interested in the average PEs over the quantum trajectories
\begin{align}
	S_q =\mathbb{E}_\mathbf{m}\left[||(|K_\mathbf{m} \psi\rangle)||^2(S_q(|\psi_\mathbf{m}\rangle))   \right]=  \mathbb{E}_\mathbf{m}\left[ \frac{1}{1-q} ||K_\mathbf{m}|\psi\rangle||^{2} \ln \left(\frac{\sum_{\vec{\sigma}} |\langle \vec{\sigma} | K_\mathbf{m} |\psi\rangle|^{2q}}{||K_\mathbf{m}|\psi\rangle||^{2q}}\right) \right],\label{eq:replicasym1}
\end{align} 
which we map below to a free energy cost of classical statistical mechanics model. Notice that here, for convenience, we consider the natural logarithm in Eq.~\eqref{eq:replicasym1} instead of the $\log_2$ used  for $d=2$. This will change the result by an overall multiplicative factor $\ln(2)$, but has the advantage of simplifying the intermediate computations.
Let us specify the notation we use in the following. The unitary gates are drawn according to the Haar distribution on $SU(d^2)$, while the measurement are chosen according to the Born-Von Neumann postulate, from the Kraus operators $\mathcal{M}_p\equiv \{1,P_1,\dots,P_d\}$, equipped with classical weights $w(1)= 1-p$, $w(P_i) = p$ which ensure that $\sum_{M\in \mathcal{M}_p} w(M) M^\dagger M =1$.
We denote the average over the unitaries and the average over the measurements $\mathbb{E}_{M\in \mathcal{M}_p}$. (Notice that these types of average are implicitly included in $\mathbb{E}_\mathbf{m}$.)
Lastly, we occasionally and conventionally  interchange the notation   $\vec{\sigma}$ with $\{\sigma\}$. 

A key remark is that the bulk of the classical statistical model is fixed by the spatiotemporal structure of the circuit, whereas the observable of interest is fully encoded in the boundary condition of the model.
As a consequence, for the problem of interest, the degrees of freedom, the geometry of the lattice and the interaction terms are the same of those presented in, \emph{e.g.}, ~\cite{jian2020measurementinduced}.

The mapping consists of two steps. First, the PEs of a single trajectory are expressed as a trace over an operator acting on $q$-copies of the system (R\'enyi replica). Then the logarithm entering the participation entropy is expanded introducing $k$ copies of the R\'enyi replicated systems. The final model involves $Q=kq+1$ layers, with the additional term coming from the circuit probability density $||K_\mathbf{m}|\psi\rangle||^2$.

We define the boundary operator $\Lambda_q$ such that
\begin{align}
	\sum_{\vec{\sigma}} |\langle \vec{\sigma} | K_\mathbf{m} |\psi\rangle|^{2q} &= \sum_{\vec{\sigma}} (\langle \vec{\sigma}| K_\mathbf{m}|\psi\rangle\langle \psi|K_\mathbf{m}^\dagger|\vec{\sigma}\rangle)^q = \sum_{\vec{\sigma}} \mathrm{Tr} (|\vec{\sigma}\rangle\langle \vec{\sigma}|K_\mathbf{m} |\psi\rangle\langle\psi| K_\mathbf{m}^\dagger |\vec{\sigma}\rangle\langle \vec{\sigma}|)^q  \nonumber \\
	& = \sum_{\vec{\sigma}} \mathrm{Tr} \left[\left(|\vec{\sigma}\rangle\langle \vec{\sigma}|K_\mathbf{m} |\psi\rangle\langle\psi| K_\mathbf{m}^\dagger |\vec{\sigma}\rangle\langle \vec{\sigma}|\right)^{\otimes q}\right] = \mathrm{Tr}  \left[\sum_{\vec{\sigma}}\left(|\vec{\sigma}\rangle\langle \vec{\sigma}|K_\mathbf{m} |\psi\rangle\langle\psi| K_\mathbf{m}^\dagger |\vec{\sigma}\rangle\langle \vec{\sigma}|\right)^{\otimes q}\right]\nonumber\\ 
	&\equiv \mathrm{Tr} \left[\Lambda_q(K_\mathbf{m}|\psi\rangle \langle \psi | K_\mathbf{m}^\dagger)^{\otimes q}\right].
\end{align}
With this notation, the average participation entropy is given by
\begin{align}
	{S_q} &= \lim_{k\to 0}\mathbb{E}_\mathbf{m} \frac{1}{k(1-q)} \left[\left(\mathrm{Tr}(\Lambda_q (K_\mathbf{m} |\psi\rangle\langle \psi| K_\mathbf{m}^\dagger)^{\otimes q})
	\right)^k - \mathrm{Tr}\left(\left( K_\mathbf{m} |\psi\rangle\langle\psi| K_\mathbf{m}^\dagger\right)^{\otimes q}\right)^{k}\right] \mathrm{Tr}(K_\mathbf{m}|\psi\rangle\langle \psi|K_\mathbf{m}^\dagger)\nonumber \\
	& =\lim_{k\to 0}\mathbb{E}_\mathbf{m}  \frac{1}{k(1-q)} \mathrm{Tr} \left[(\Lambda_q^{\otimes k}-1)\left( K_\mathbf{m} |\psi\rangle\langle\psi| K_\mathbf{m}^\dagger\right)^{\otimes kq +1}\right]\equiv \frac{q}{1-q} \lim_{k\to 0} \frac{ \mathcal{Z}_\Lambda - \mathcal{Z}_0}{Q-1}.
	\label{eq:rep2}
\end{align}
In the above expression we have defined the partition functions
\begin{align}
	\mathcal{Z}_\Lambda &= \mathbb{E}_\mathbf{m} \mathrm{Tr}(\Lambda_q^{\otimes m} (K_\mathbf{m}|\psi\rangle\langle \psi| K_\mathbf{m}^\dagger)^{\otimes Q}),\label{eq:zpart1}\\
	\mathcal{Z}_0 &= \mathbb{E}_\mathbf{m} \mathrm{Tr}((K_\mathbf{m}|\psi\rangle\langle \psi| K_\mathbf{m}^\dagger)^{\otimes Q}).\label{eq:zpart2}
\end{align}
Pictorially, we can introduce the layer representation of a circuit realization
\begin{equation}
    \dia{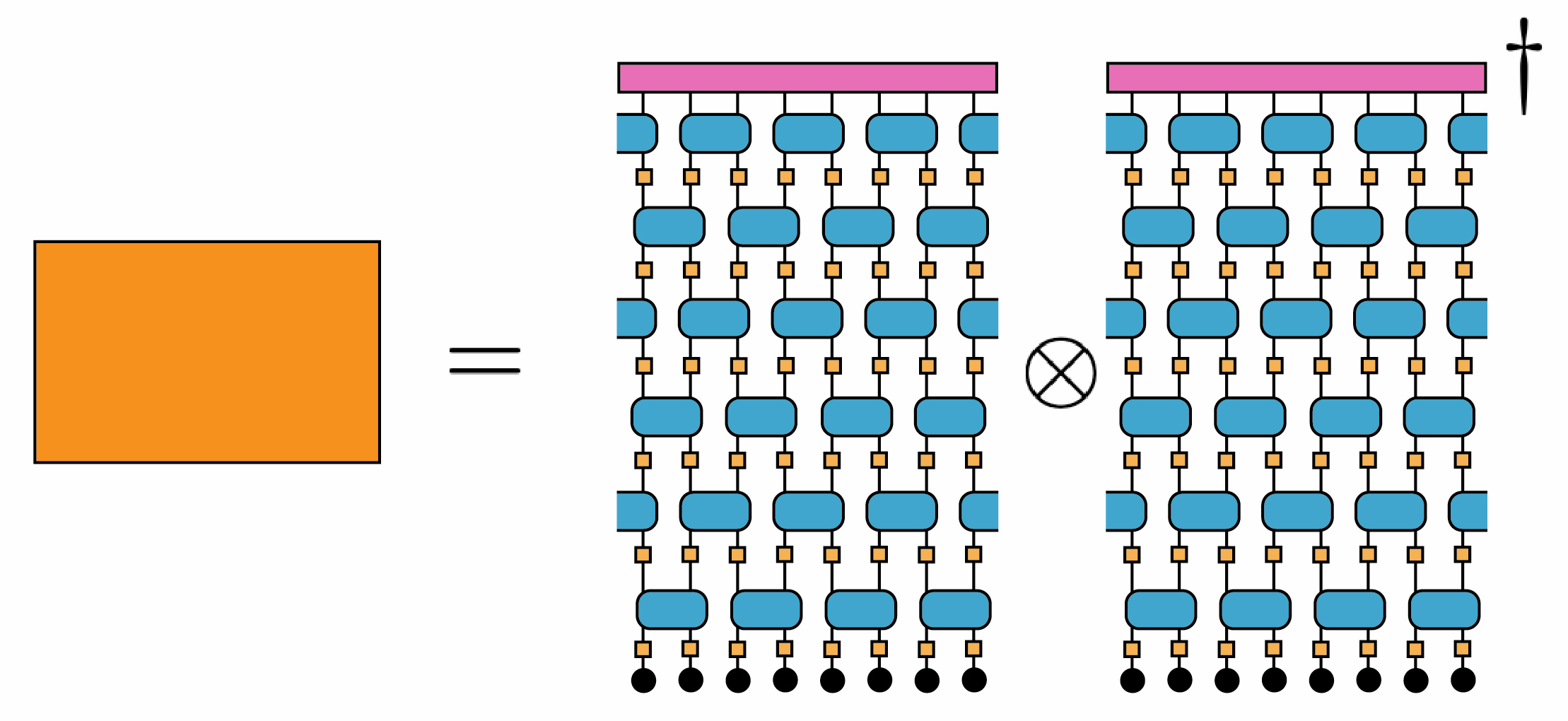}{-64}{120}\label{eq:zione}
\end{equation}
which recasts Eqs.~\eqref{eq:zpart1} and ~\eqref{eq:zpart2} to \begin{equation}
    \dia{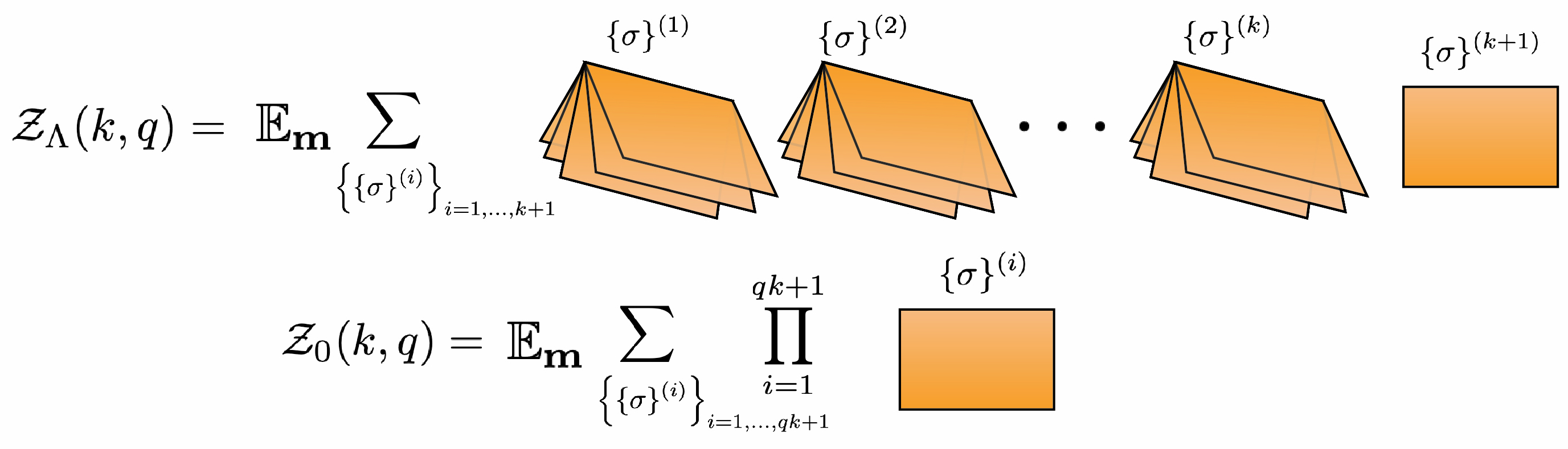}{-64}{120}\label{eq:diagram2}.
\end{equation}

Since $\mathcal{Z}_\Lambda = \mathcal{Z}_0 = 1$ in the replica limit $k\to 0$ ($Q\to 1$), Eq.~\eqref{eq:rep2} becomes a free-energy cost associated with change of the boundary condition
\begin{equation}
    S_q =  F_\Lambda - F_0 \equiv \lim_{k\to 0} \frac{\ln(\mathcal{Z}_\Lambda/\mathcal{Z}_0)}{k(1-q)},\label{eq:freeeneg}
\end{equation}
where  $F_\alpha = \lim_k (\ln \mathcal{Z}_\alpha)/(k(1-q))$ are the free energies for the boundary condition $\alpha=0,\Lambda$.

Since both measurement and unitary gates are independently and randomly distributed, we factorize the average $\mathbb{E}_\mathbf{m}$.
For the unitary gates we have
\begin{equation}
    \mathbb{E}_{U}[U_{i,i+1}^{\otimes Q} U_{i,i+1}^{\dagger \otimes Q}] =  \sum_{s,r\in\mathbb{S}_Q} \mathrm{Wg}_{d^2}(s r^{-1}) \chi_i(s) \chi_{i+1}(s) \chi_{i}(r)\chi_{i+1}(r) =\dia{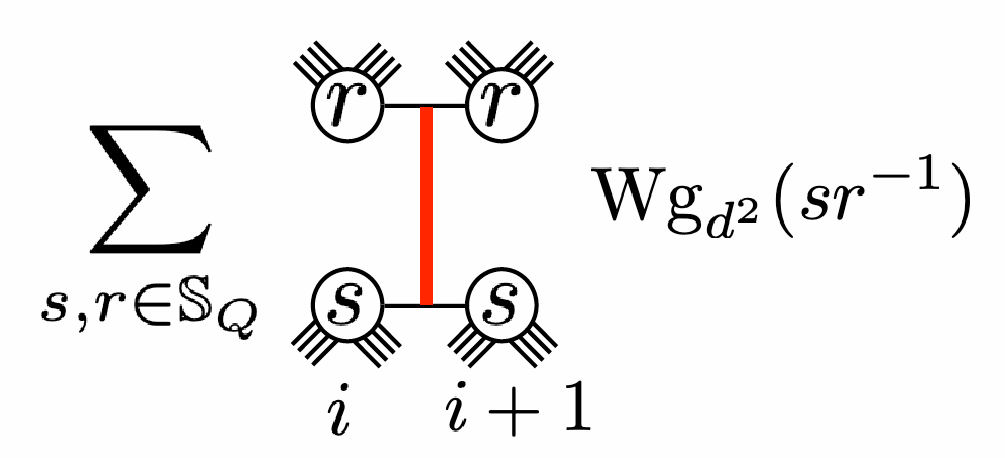}{-38}{68},\label{eq:unit_rep}
\end{equation}
where $\mathrm{Wg}_D(s)$ are the Weingarten symbols, and $s,r$ are permutations over the symmetry group $\mathbb{S}_Q$~\cite{jian2020measurementinduced}.
The operators $\chi_s$ have matrix elements
\begin{equation}
    \chi_{\{\sigma\},\{\tau\}}(s) \equiv \langle \sigma^{(1)}\sigma^{(2)}\cdots \sigma^{(Q)}|\chi(s)|\tau^{(1)}\tau^{(2)}\cdots \tau^{(Q)}\rangle = \prod_{r=1}^Q \delta(\sigma^{(r)},\tau^{(s(r))}).\label{eq:matelem}
\end{equation}

The average over the measurements can be performed in a similar fashion. It is convenient to consider the measurements contracted with the operators in Eq.~\eqref{eq:unit_rep}, leading to
\begin{equation}
    \dia{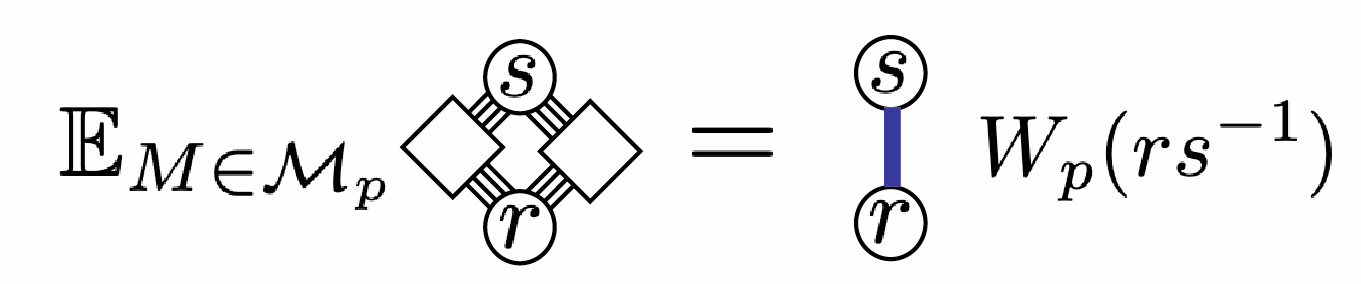}{-25}{35}
\end{equation}
where the square indicates $M^{\otimes Q}$, and  $W_p(s) = (1-p) d^{\mathcal{C}(s; \mathbb{S}_Q)}+pd$ and $\mathcal{C}(s; \mathbb{S}_Q)$ is the number of cycles in the permutation $s$ in $\mathcal{S}_Q$.
The action on the initial (product) state $|\psi\rangle= \otimes_{a=1}^L |e\rangle_a$ is given by
\begin{equation}
    \dia{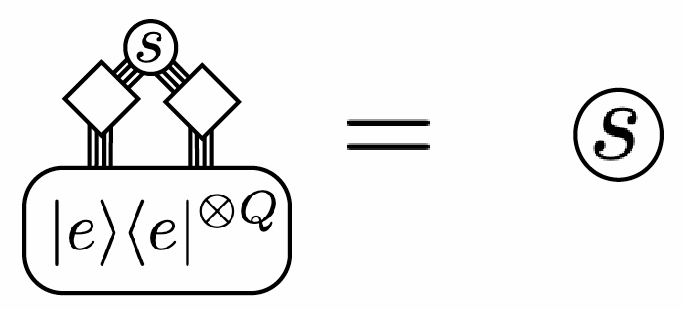}{-10}{35}\
\end{equation}
Instead, the boundary condition $\alpha$ is fixed by the operator acting on the top layer of the circuit
\begin{equation}
    \dia{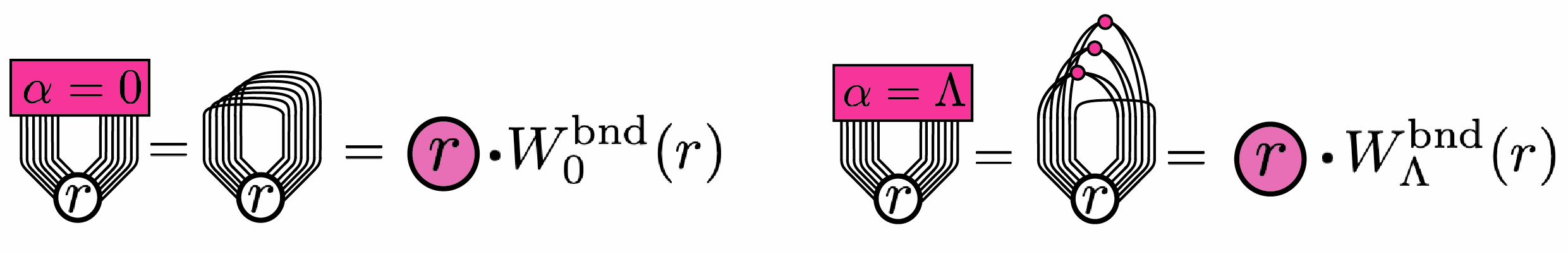}{-10}{45}\label{mamma}
\end{equation}
where the contraction signals a sum over the associated degrees of freedom. In particular, we have
\begin{equation}
    W_\alpha^\mathrm{bnd}(s) = \sum_{\sigma_1,\sigma_2\dots,\sigma_Q=1}^d \chi_{\{\sigma\},\{\sigma\}}\times \begin{cases}
        1,&\alpha=0\\
        \prod_{r=0}^{(Q-1)/k-1} \prod_{a=1,\dots,q-1} \delta_{\sigma_{kr+1},\sigma_{kr+1+a}}, &\alpha=\Lambda
    \end{cases}
\end{equation}
where the delta functions in the $\Lambda$ (pictorially represented in the magenta dots in Eq.~\eqref{mamma}) impose the ``book'' structure on the $k$ replicae (see Eq.~\eqref{eq:diagram2}).

After the average is performed, the partition function becomes a classical spin model on an anisotropic honeycomb lattice $\Xi$
\begin{equation}
    \dia{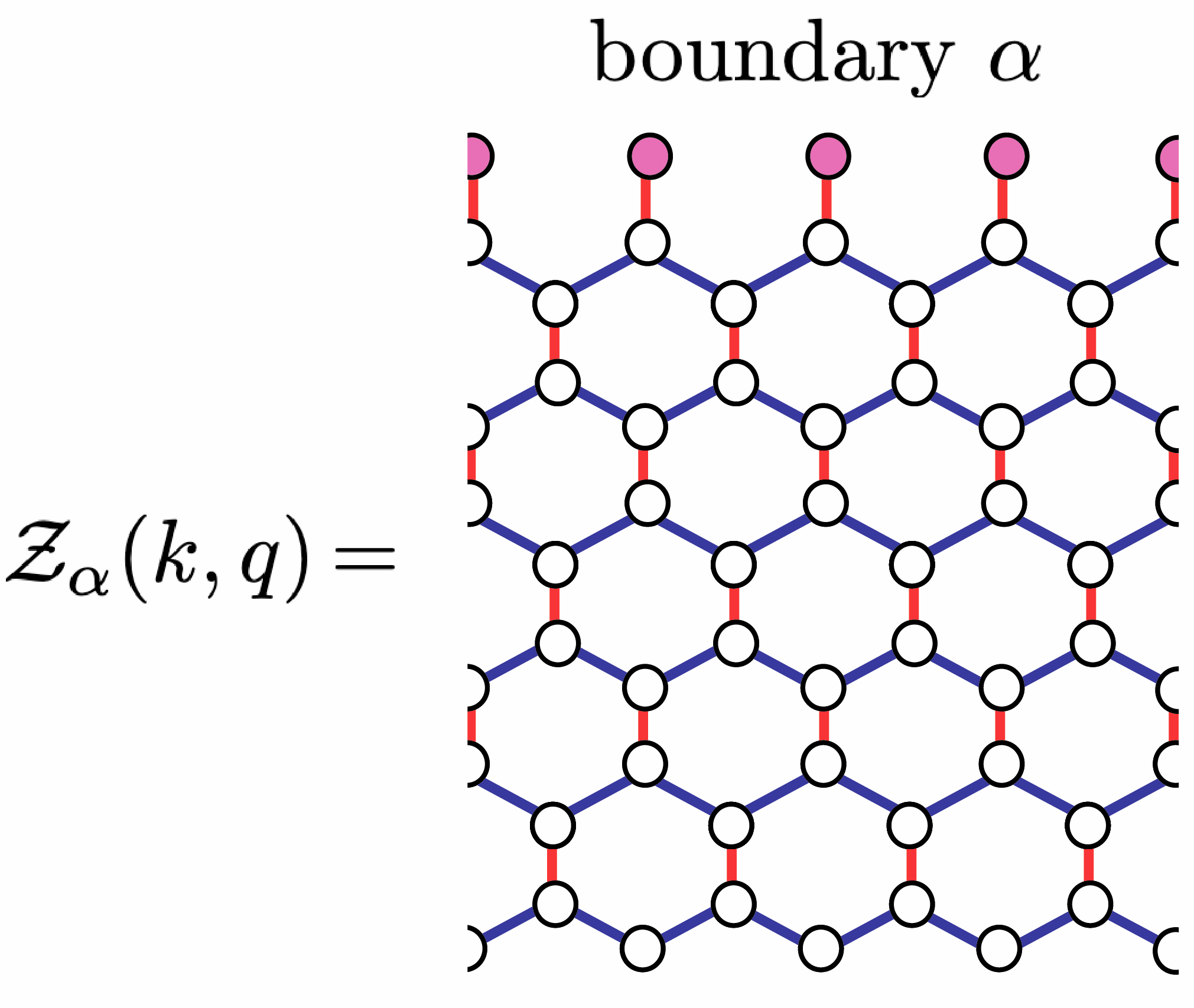}{-60}{140} = \sum_{\{s_i\in \mathbb{S}_Q\}} \prod_{\langle ij\rangle \in \color{blue}{\Xi}}W_p(s_i s^{-1}_j) \prod_{\langle ij\rangle \in \color{red}{\Xi}}\mathrm{Wg}(s_i s^{-1}_j) \prod_{i\in \color{magenta}{\partial \Xi}}W_\alpha^{\mathrm{bnd}}(s_i).\label{eq:anis}
\end{equation}
In Eq.~\eqref{eq:anis}, the spins are elements of the symmetric group $\mathbb{S}_Q$, and we have introduced the notation $\color{red}{\Xi}$ to define the red links, $\color{blue}{\Xi}$ the blue links, and $\color{magenta}{\partial\Xi}$ the boundary sites.
The lattice contains $N=L (t-1)/2$ spins for a qudit chain of length $L$ and for a total time $t\gg 1$.

\subsection{Large $d$ limit}
\label{sec:rep2}
The expression Eq.~\eqref{eq:anis} involves no approximation, and is valid for any $d$. However, the full computation of the partition function $\mathcal{Z}_\alpha(k,q)$ is involved, due to frustration effects in its weights. To obtain analytic insights we therefore consider the limit of large qudit dimension $d\gg 1$. At leading order we have
\begin{align}
    \mathrm{Wg}(s) &= \delta_{s} d^{-2Q}(1+ O(d^{-2})),\qquad W_p(s) = d^Q((1-p)\delta_s + p d^{1-Q})+ O(d^{Q-1}),\\
    W_0^\mathrm{bnd}(s) & = \delta_s d^Q (1+O(d^{-1})),\label{eq:conts1}\qquad W_\Lambda^\mathrm{bnd}(s) = \Delta_s(\mathbb{S}_q^{\otimes k}\otimes 1) d^{k+1}(1+O(d^{-1})),
\end{align}
where we have introduced the function
\begin{equation}
    \delta_s = \begin{cases} 1 &\text{if } s=\mathbf{1},\\
    0 & \text{otherwise},\end{cases}
\end{equation}
and 
\begin{equation}
    \Delta_s(G) = \begin{cases} 1 &\text{if } s\in G,\\
    0 & \text{otherwise},\end{cases}
\end{equation}
and $\mathbf{1}$ is the identity permutation.
The leading order of $W^\mathrm{bnd}_\Lambda(s)$ is a consequence of the permutation invariance of each book page. The sub-leading corrections in $d$ 
arise when larger groups are considered. For instance, $\mathbb{S}_{2q} \otimes \mathbb{S}_q^{\otimes (k-1)}\otimes 1$ would arise when the first two book share the same boundary state, which is equivalent to consider them merged in a single book with $2q$ pages. 
Another way to obtain a sub-leading in $d$ correction is to add the $Q$-th replica to the $k$-th book (hence forming a new book of $q+1$ pages), which gives rise to group $\mathbb{S}_q^{\otimes (k-1)}\otimes \mathbb{S}_{q+1}$.

Hence, up to corrections in $d^{-2}$, the honeycomb lattice reduces to a square one
\begin{equation}
    \dia{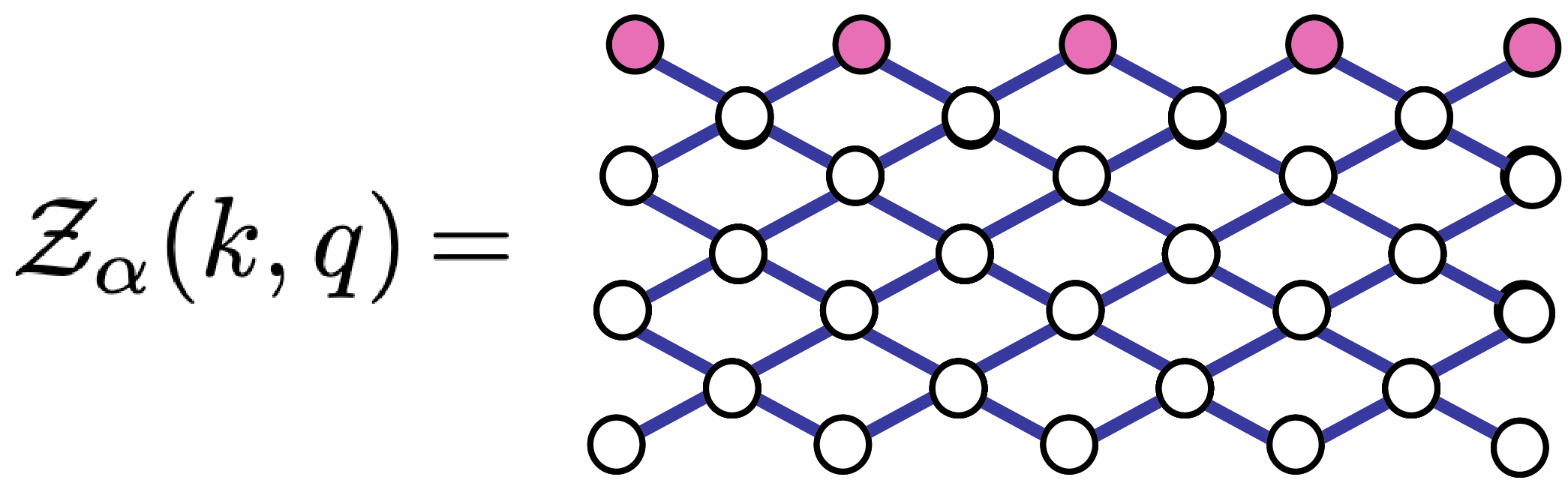}{-30}{50},\label{eq:square}
\end{equation}
with $2N$ links.

The expression for the partition function becomes 
\begin{equation}
    \mathcal{Z}_\alpha = \sum_{\{s_i \in \mathbb{S}_Q\}} e^{-\beta \sum_{\langle i,j\rangle} v_{i,j}}\left(\prod_{r\in \partial\Xi} W^\mathrm{bnd}_\alpha(s_r)\right) d^{-QL},
\end{equation}
where the Boltzmann weights can be easily expressed from $W_p(s)$ and $W_\alpha^\mathrm{bnd}(s)$, and are given by
\begin{equation}
    \beta = |\ln p|,\qquad v_{i,j} = (1-\delta_{s_i s_{j}^{-1}}) + (1-Q)\ln d  + O(d^{-1}).
\end{equation}
(Notice that here we considered $Q\ge 1$).
This expression shows that $|\ln p|$ plays the role of an inverse temperature.
Hence, at the leading order the model is a $Q!$-state Potts model defined on the structure Eq.~\eqref{eq:square}. Neglecting the constant energy term, which simplifies in the ratio Eq.~\eqref{eq:freeeneg}, we have
\begin{align}
    \mathcal{Z}_0(k,q)& = \sum^\mathrm{bulk}_{\{s_i \in \mathbb{S}_Q\}} e^{-\beta \sum_{\langle i,j\rangle} v_{i,j} } \\
    \mathcal{Z}_\Lambda(k,q)& = \sum^\mathrm{bulk}_{\{s_i \in \mathbb{S}_Q\}} \sum^\mathrm{bnd}_{\{s_i \in \mathbb{S}_q^{\otimes k}\otimes 1\}}e^{-\beta \sum_{\langle i,j\rangle} v_{i,j} } d^{(k+1-Q)L}.
\end{align}

\subsection{Cluster expansion at $p\to 0$ and $p\to 1$}
\label{sec:rep3}
We conclude this section by considering the limit of low and high measurement rate, which correspond, respectively, to the low $(\beta\to\infty)$ and high $(\beta \to 0)$ temperature limits. 

At low temperature, the weights are zero, except when the $v_{i,j}=0$. Hence, we can approximate 
\begin{equation}
    e^{-\beta v_{i,j} }\overset{p\simeq 0}\simeq ((1-p)\delta_{s_i s_j^{-1}}+p).
\end{equation}
We expand the partition functions in a series in $p$: $\mathcal{Z}_\alpha(k,q)=\sum_{n=0}^{2N} \frac{p^n}{n!} \mathcal{Z}_\alpha^{(n)}(k,q)$. A simple combinatoric computation gives
\begin{equation}
    \mathcal{Z}_0^{(0)}(k,q) = 1,\qquad \mathcal{Z}_\Lambda^{(0)}(k,q)= \Gamma(1+q)^{k+1} d^{(k+1-Q)L}. 
\end{equation}
Here the combinatoric constant in $\mathcal{Z}_\Lambda(k,q)$ comes from the number of permutations satisfying the constraint Eq.~\eqref{eq:conts1} (i.e. $s \in \mathbb{S}_q^{\otimes k}\otimes 1$) and from the condition of having the $\delta$-function on each link.
We note that this condition enforces a ferromagnetic-like phase, with the bulk degrees of freedom fixed by the boundary ones. 
The first sub-leading correction is second-order in $p$ and is given by
\begin{equation}
    \mathcal{Z}^{(2)}_0(k,q) = L,\qquad \mathcal{Z}^{(2)}_\Lambda(k,q) = L \Gamma(1+q)^{2 k}.
\end{equation}
Therefore, in the limit $p\to 0$ we have
\begin{equation}
    S_q = \lim_{k\to 0} \frac{1}{k(1-q)}\log_2 \left(\frac{\mathcal{Z}^{(0)}_\Lambda + \frac{p^2}{2}\mathcal{Z}^{(2)}_\Lambda }{\mathcal{Z}^{(0)}_0 + \frac{p^2}{2}\mathcal{Z}^{(2)}_0}\right)= L\left(\log_2(d)-\frac{p^2}{2}\frac{\log_2 \Gamma (1+q)}{q-1}\right) + \frac{\log_2 \Gamma(1+q)}{1-q},
\end{equation}
where we reabsorbed the $\ln 2$ from Eq.~\eqref{eq:replicasym1} for comparison with the Main Text.
Importantly, we note that the value of the sub-leading term $c_q= \log_2(\Gamma(1+q))/(1-q)$ is not affected by the increase of the measurement rate $p$ (within the range of validity of the second order approximation in $p$). This 
is consistent with
the numerical evidence presented in the Main Text, supporting $c_q \neq 0$ within the error-correcting phase.

In the opposite limit of infinite temperature ($p\to 1$), it is convenient to express the Boltzmann weights as
\begin{equation}
    e^{-\beta v_{i,j} }\overset{p\simeq 1}\simeq (1+(1-p) f_{i,j}),\qquad \text{with } f_{i,j} = (\delta_{s_i s_j^{-1}} -1), 
\end{equation}
an expand the partition functions in a series in $(1-p)$ as $\mathcal{Z}_\alpha(k,q) = \sum_{n=0}^{2N} \frac{(1-p)^n}{n!} \mathcal{Z}^{(n)}_\alpha(k,q)$. In this limit, the spin system is in a paramagnetic phase: the permutations on different lattice sites are independent of each other. Hence, the partition function is approximated by an equal weight sum over all the possible degrees of freedom.
A combinatorical computation gives
\begin{equation}
    \mathcal{Z}_0^{(0)}(k,q) = \Gamma(1+Q)^{N},\qquad \mathcal{Z}_\Lambda^{(0)}(k,q) = \Gamma(1+Q)^{N} \Gamma(1+q)^{k L/2}.
\end{equation}
Similarly, it is possible to compute the first sub-leading corrections, which are order $O((1-p)^2)$
\begin{align}
    \mathcal{Z}_0^{(2)}(k,q) & = \Gamma(1+Q)^N ( 2N^2 + N(N-1) \Gamma(1+Q)^{-2} - 4 N^2 \Gamma(1+Q)^{-1})\\
    \mathcal{Z}_\Lambda^{(2)}(k,q) & = \Gamma(1+q)^{kL/2}\Gamma(1+Q)^N ( L \Gamma(1+q)^{-1} \Gamma(1+Q)^{-1} + \nonumber \\&\qquad \qquad(N(N-1)-L) \Gamma(1+Q)^{-2} + 2 N^2 - 2 N^2 \Gamma(1+Q)^{-1}).
\end{align}
Hence, at high temperature, the PEs reads
\begin{equation}
    S_q = L\left(\log_2 d + \frac{1}{2} \frac{\log_2 \Gamma(1+q)}{1-q} + \frac{ (1-p)^2}{2(q-1)}\log_2 \Gamma(1+q)\right),
\end{equation}
where, again, we reabsorbed the $\ln 2$ from Eq.~\eqref{eq:replicasym1} for comparison with the Main Text.
Our computation is consistent with the numerical analysis for $d=2$, and in particular support the stability of $c_q=0$ within the quantum Zeno phase $p>p_c$.

\section{S6 Additional Numerical results}
In order to test the universality of the behavior of the sub-leading term $c_q$ in scaling of PEs, in this section we numerically investigate three additional setups: (i) rank-2 measurement stabilizer circuits, (ii) $d=3$ qudit Haar circuits, and (iii) Floquet quatum circuits with weak measurements. 
Throughout this section, we consider the PEs in the $Z$-basis computed after the layer of unitary gates.

\begin{figure}[th]
    \centering
    \includegraphics[width=.7\columnwidth]{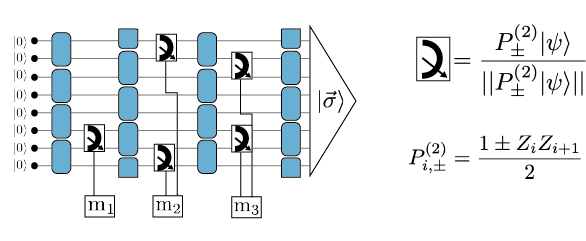}
    \caption{Scheme for the stabilizer circuits with the rank-2 measurements $P^{(2)}_\pm$. The unitary gates $U_{i,i+1}$ are drawn from the Clifford group  as in Fig.~1 of the Main Text.}
    \label{fig:zzscheme}
\end{figure}

\subsection{Rank-2 measurements stabilizer circuits}
We consider the stabilizer circuit in Fig.~\ref{fig:zzscheme}. The difference  with respect to the setup in Fig.~1 of the Main Text is the substitution of the rank-1 measurement $P^{(1)}_{i,\pm} \equiv P_{i,\pm} = (1+Z_i)/2$ by rank-2 measurements $P^{(2)}_{i,\pm} = (1\pm Z_i Z_{i+1})/2$. These operators project onto the following Bell states 
\begin{align}
    P^{(2)}_{i,+}&= \frac{1}{2}(|00\rangle + |11\rangle)(\langle 00 | + \langle 11|)\;,\nonumber \\
    P^{(2)}_{i,-}&= \frac{1}{2}(|01\rangle + |10\rangle)(\langle 01 | + \langle 10|)\;.\label{eq:bellissimo}
\end{align}
We note that the rank-2 measurements have less disentangling power compared to the rank-1 ones, as the residual entanglement of the Bell states in Eq.~\eqref{eq:bellissimo} is $\ln 2$. Therefore the critical point of this system is expected to occur to higher measurement rates.

This setup has been previously considered in Ref.~\cite{li2018quantum}. We note that the exponent $\nu$ estimated in that paper ($\nu \simeq 1.75$) does not fit later and more accurate studies for the stabilizer circuits with rank-1 measurements ($\nu=1.30(5)$ ~\cite{gullans2020scalable}). (The reason for this discrepancy is due to a smaller accuracy of the entanglement entropy collapses considered in Ref.~\cite{li2018quantum} in comparison to the entanglement measures considered in ~\cite{li2019measurementdriven,gullans2020scalable}.)
Thus, our analysis has an additional benefit. It extracts a precise location of the critical point and an accurate value the correlation length exponent which have been not yet reported for stabilizer circuits with rank-2 measurements.

\begin{figure}[h]
    \centering
    \includegraphics[width=0.8\columnwidth]{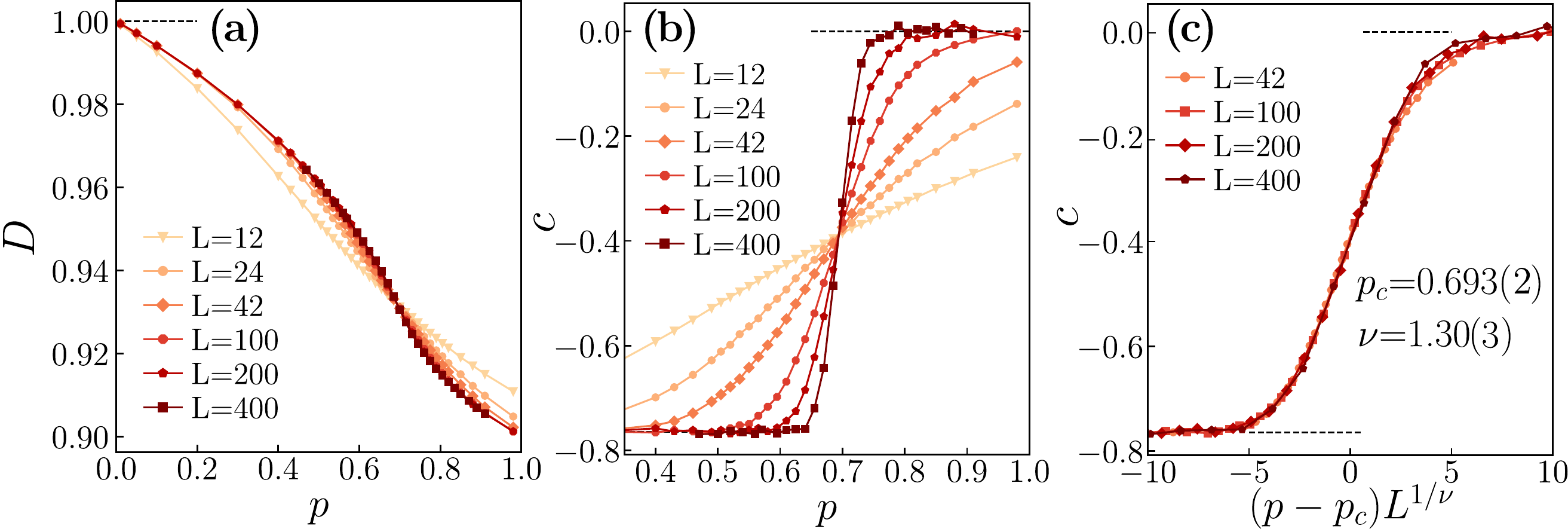}
    \caption{Results for the stabilizer circuits with rank-2 measurements. (a) Fractal dimension for the system for various $p$ and various system sizes. (b) The sub-leading term $c$ for various $p$ and various system sizes $L$. (c) Scaling collapse of $c$.  }
    \label{fig:zz}
\end{figure}

The numerical results, averaged over more than 20000 circuit realizations, are given in Fig.~\ref{fig:zz}. We see that, similarly to the rank-1 case, the fractal dimension $D$ converges at large system sizes, to a universal curve, whereas the sub-leading term $c$ approaches a step-function for increasing system size $L$, with $c=c^\mathrm{stab}\simeq -0.76$ for $p<p_c$ and $c=0$ for $p>p_c$. The estimated critical point is $p_c=0.693(2)$ and the correlation length critical exponent is $\nu=1.30(3)$. 

We remark that both the value of the sub-leading term $c$ in the error correcting phase, and of the exponent $\nu$ at the critical point match those for the rank-1 measurements discussed in the Main Text. This  shows that the  stabilizer circuits with rank-1 and rank-2 measurements lie in the same universality class, and that $c$, as a property of the phase, only depends on gross features of the model (such as the state being a stabilizer or not).

\subsection{$d=3$ qudit Haar circuits}
In this subsection we consider the circuits with spatiotemporal structure shown in the Fig.~1 of the Main Text, but with qudit dimension $d=3$, with $U_{i,i+1}\in SU(9)$ drawn with Haar measure and with $P_{-1,0,+1}$ projecting onto the eigenvectors  of 
\begin{equation}
    Z \equiv \begin{pmatrix} -1 & 0 & 0 \\ 
    0 & 0 & 0 \\
    0 & 0 & 1\end{pmatrix}.
\end{equation}

To the best of our knowledge this setup has not been previously discussed in the literature, therefore it constitutes a novel test for the universality of Haar circuits. 
Importantly, it is expected that the qudit dimension  $d$  is a relevant perturbation for the MIPT in random Haar circuits, since the limit $d=\infty$ differs from the finite $d$~\cite{skinner2019measurementinduced}. 

\begin{figure}[h]
    \centering
    \includegraphics[width=\columnwidth]{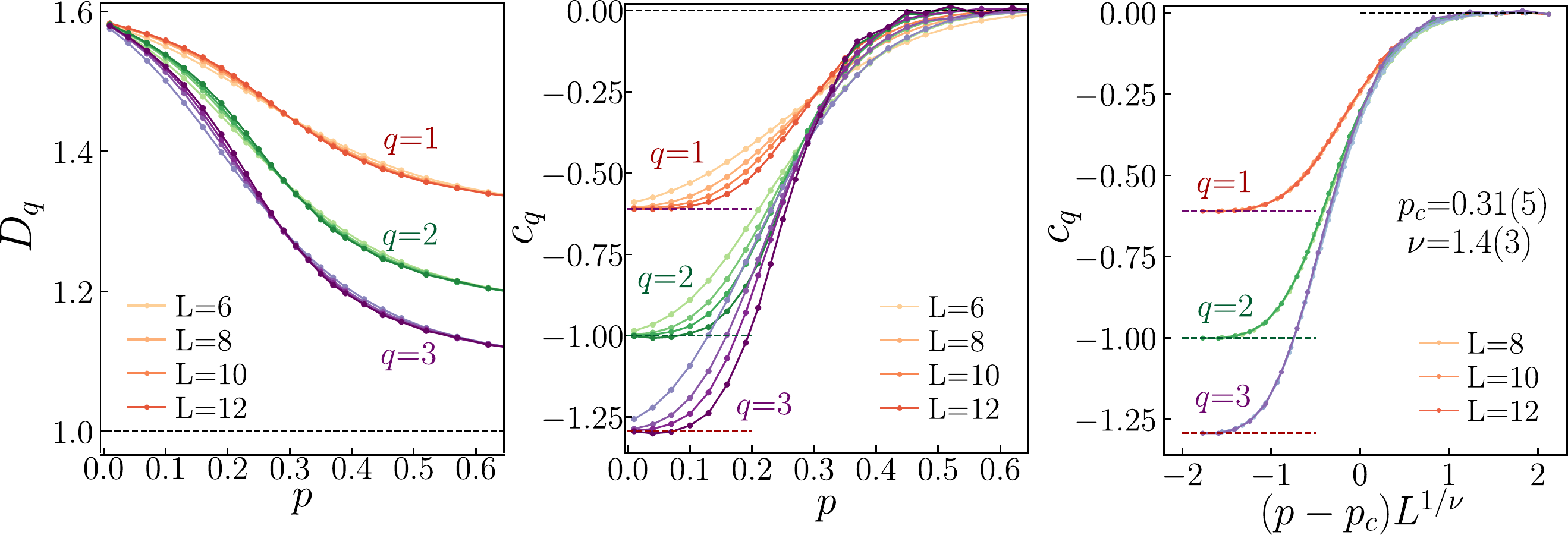}
    \caption{\label{fig:d3} Numerical analysis for the Haar random circuits with $d=3$. (a) Multifractal dimension $D_q$ for different $p$ and for different system sizes $L$. (b) Sub-leading term $c_q$ for different values of $p$ and for different values of $L$. (c) Scaling collapse for $c_q$.}
\end{figure}

In our numerical simulations we consider systems of up to $L=16$ sites.
This corresponds to Hilbert space dimension $\mathcal N = 3^{16}= 43046721$, larger that the largest Hilbert space dimension for $d=2$ case considered by us or in \cite{zabalo2020critical}.
Our findings are gathered in Fig.~\ref{fig:d3}. The results are averaged over more than $2000$ disorder realizations.
While, we  believe that the system size is insufficient to extract a precise position of the critical point and of the exponent $\nu$, we obtain a good collapse of the data for the sub-leading term $c_q$ with $p_c=0.31(5)$ and $\nu=1.4(3)$. We note that the exponent $\nu$ is compatible with the numerical result for $d=2$ given in the Main Text.  

Furthermore, we note that the value of $c_q$ in the error correcting phase matches those estimated for $d=2$ and for ${d \to \infty}$ in our paper: $c_q=(1-q)^{-1}\log_2 \Gamma(1+q)$. Hence, we conclude that also for Haar circuits the characterization of the phase through $c_q$ depends only on the gross features of the circuits. This value of $c_q$ precisely
coincides with the result for Gaussian Unitary Ensemble (GUE) of random matrices \cite{Backer19}, appropriate for systems with broken time reversal symmetry. It is, however, important to note that the GUE result $D_q=1$ and $c_q=(1-q)^{-1}\log_2 \Gamma(1+q)$ describes our systems only at $p \to 0$. For $p>0$, the wave-functions in random circuits with measurements become, in general, multifractal ($D_q<1$) but the sub-leading term persist to be equal to $c_q=(1-q)^{-1}\log_2 \Gamma(1+q)$ in the whole error-correcting phase $p<p_c$.

Since there are no works in the literature discussing the Haar circuits with $d=3$, it is interesting to compare the results obtained from the analysis of PEs with the results obtained through entanglement measures. Specifically, we consider the bipartite mutual information $I_2$ and the tripartite mutual information $I_3$, which are considered as precise observables to locate the critical point in random circuits~\cite{zabalo2020critical}. 

\begin{figure}[h]
    \centering
    \includegraphics[width=0.67\columnwidth]{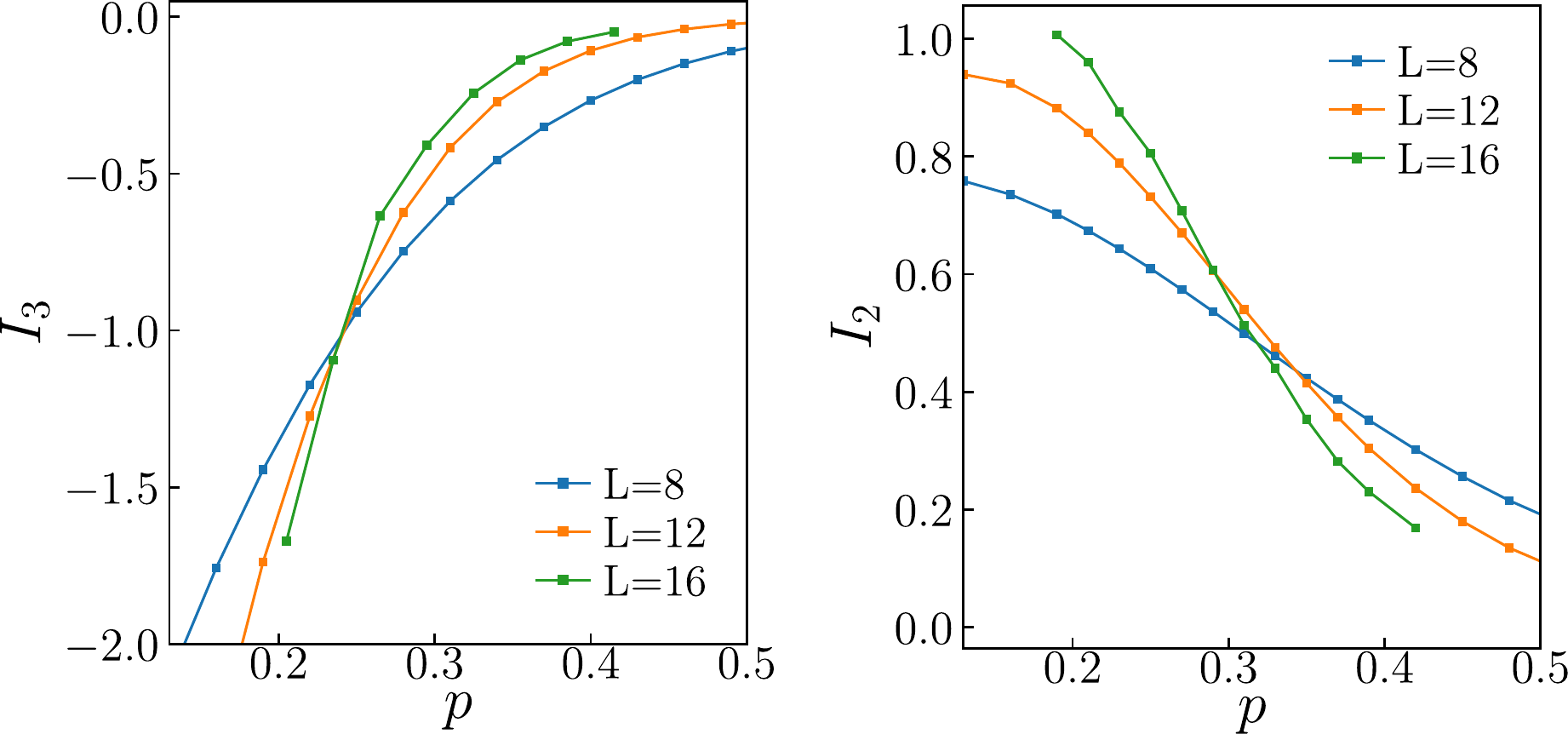}
    \caption{
    Bipartite (right panel) and tripartite mutual information (left panel) for $d=3$ Haar circuits. The plots  show results for  different system sizes $L$ and different measurement rates $p$.
    \label{fig:d3i3}}
\end{figure}

These quantities are both defined in terms of the entanglement entropy. Given a state $|\psi\rangle $ and a bipartition  $M\cup N$, the entanglement entropy between the subsystems $M$ and $N$ is given by
\begin{equation}
    S(M) = S(N) = -\mathrm{tr}\rho_M\ln \rho_M\;,\qquad \rho_M = \mathrm{tr}_N|\psi\rangle\langle \psi|.
    \label{eq:ent}
\end{equation}
We consider a partition of our system (with periodic boundary conditions) into four adjacent (and not overlapping) subsystems $A, B, C, D$, each of length $L/4$, so that the subsystems $A$ and $C$ are antipodal. We calculate the bipartite quantum mutual information between subsystems $A$ and $C$:
\begin{equation}
    I_2 = S(A) + S(C) - S(A\cup C)\;,\label{eq:i2}
\end{equation}
where $S(A)$ is the entanglement entropy \eqref{eq:ent} with $M=A$ and $N=B \cup C \cup D$. 
The tripartite mutual information is given by
\begin{equation}
    I_3 = S(A) + S(B) + S(C) - S(A\cup B) - S(A\cup C) - S(B\cup C) + S(A\cup B\cup C)\;.\label{eq:i3}
\end{equation}
We compute Eq.~\eqref{eq:i2} and Eq.~\eqref{eq:i3} after the unitary gate layer, and consider the average value over the trajectories. Our numerical results are given in Fig.~\ref{fig:d3i3}, results are averaged over more than 400 disorder realizations. 
For the considered partitions, the critical point is identified by a crossing point~\cite{zabalo2020critical}. (Differently from the choice given in Ref.~\cite{li2019measurementdriven}, where the transition is identified by a peak. The different behavior stems from the scaling of the subsystems size with the total system size.)
From the data, we find that the crossing point for the bipartite mutual information is shifting toward smaller values of $p$ with increasing system size $L$, suggesting large finite size effects. On the other hand, the tripartite mutual information exhibits a crossing point in the vicinity of $p\simeq 0.24$, which is roughly compatible with our estimate through the PEs. However, the limited sizes do not allow for a proper finite size scaling. Hence these results serve only as a consistency check.

\begin{figure}[h]
    \centering
    \includegraphics[width=0.67\columnwidth]{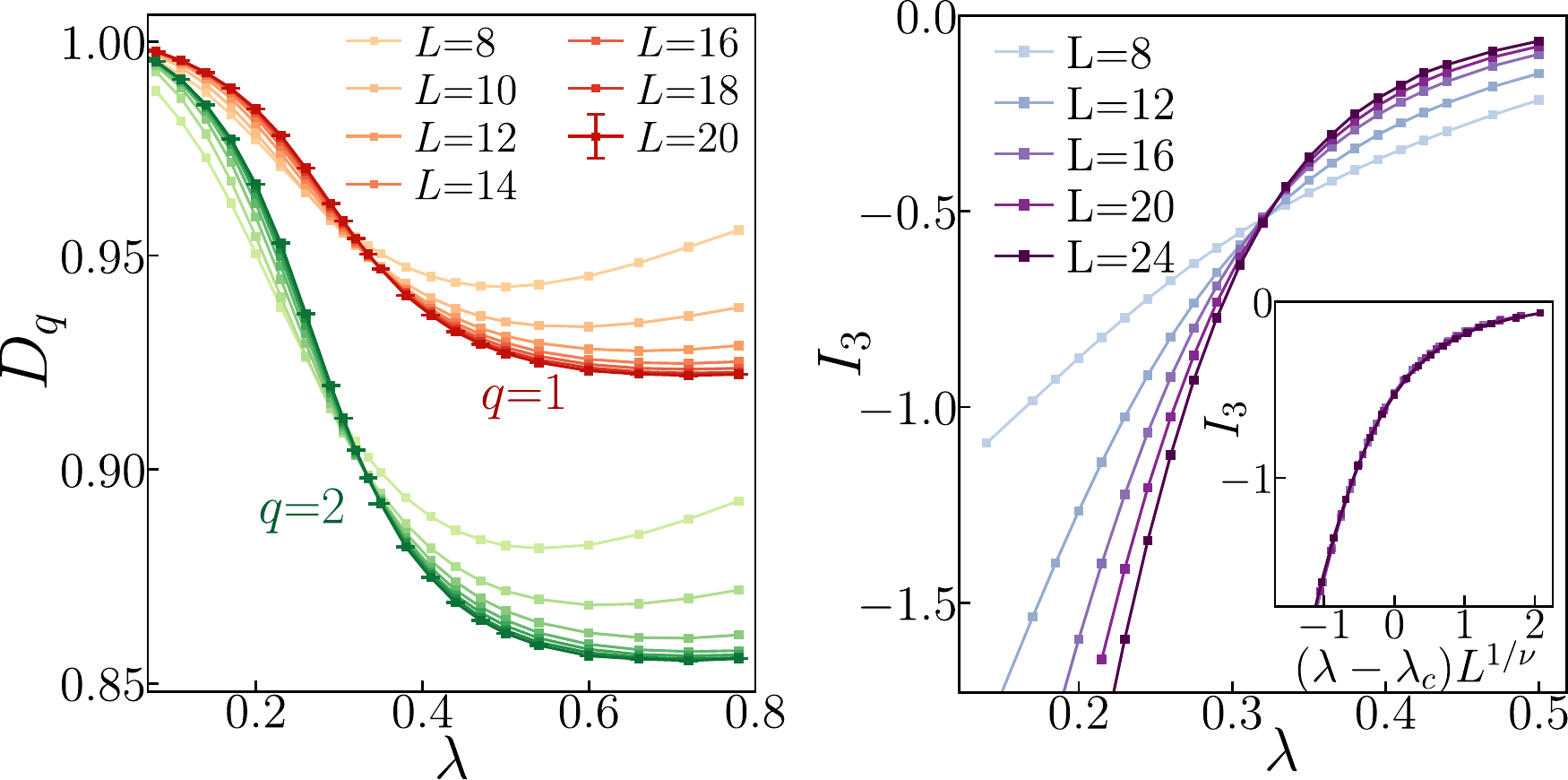}
    \caption{Results for the  multifractal dimension (left panel) and for the tripartite mutual information (right panel) for the Floquet quantum circuits with weak measurements. The inset show the scaling collapse for $I_3$, with  $\lambda_c = 0.320(8)$ and $\nu=1.3(1)$ (here the measurement rate $p=1$ is fixed). }
    \label{fig:floqqq}
\end{figure}

\subsection{Floquet quantum circuits}
Lastly, we consider Floquet quantum circuits with weak measurements. This framework provides two tests. On one hand, we see if the Haar circuits are representative of generic 
quantum evolution. On the other hand, we also test if weak measurements change the physics of the system. This framework has been previously considered in Ref.~\cite{li2019measurementdriven}; however our numerical analysis gives better estimates of the position of the critical point and of the exponent $\nu$, as we extend the numerical simulations over larger system sizes and consider finer observables for the transition.

The circuit has a similar architecture to that given in Fig.~1 of the Main Text, but here the layer of two-qubit unitary gates is replaced by the Floquet operator
\begin{equation}
    U_{F} = e^{-i \sum_j X_j} e^{ -i \sum_j ( Z_j Z_{j+1}  + Z_j ) } 
\end{equation}
while the projections $P_\pm$ are substituted with $M_{\pm}=(1\pm \lambda Z_j)/(2+2\lambda^2)$ which act on each qubit (measurement rate $p=1$) with strength determined by the parameter $\lambda$. 

We consider both the PEs as well as the tripartite mutual information \eqref{eq:i3} with the same partition of the system as for the Haar circuits. Our results, averaged over more than 1000 circuit realizations, are given in Fig.~\ref{fig:floqqq} and in Fig.~\ref{fig:3} 
of the Main Text.

The estimates for the critical measurement strength $\lambda_c$ and for the  critical exponent $\nu$ both for the sub-leading term $c_q$ and for the tripartite quantum mutual information are compatible and are given by $\lambda_c=0.320(8)$ and $\nu = 1.3(1)$.
Therefore, we conclude the Floquet circuits with weak measurements also belong to the same universality class of Haar circuits with $d=2$ qubits and projective measurements. Furthermore, we observe that in the error correcting phase, for $\lambda < \lambda_c$, the sub-leading terms in scaling of PEs are given by  $c_q=(1-q)^{-1}\log_2 \Gamma(1+q)$, and are the same as for the Haar random circuits. The value of $c_q$ mathches the GUE prediction (note that the time reversal symmetry is broken in the Floquet circuits by the presence of measurements). This suggests that $c_q$ plays the same role in this phase as for the Haar error correcting phase.

\section{S7 Experimental proposal on quantum processors}

In this section we simulate an experiment that observes the change of the sub-leading term $c_q$ in system size scaling of PEs across MIPT. This allows us to establish the experimental relevance of our findings by arguing that systems of size $L=10$ seem to be well in reach of our method on superconducting quantum processors (this system size should be compared with recent experiment performed in systems of up to $L=8$ sites \cite{noel2021observation}). The following analysis demonstrates also that the experimental resources needed to observe MIPT through the PEs scale exponentially with the size of the system.

In numerical calculations, the many-body wave function $\psi(\sigma) = \langle \sigma |\psi \rangle$ of state $\ket{\psi}$ in many-body basis $\{ \ket{\sigma} \}$ is directly stored in the memory of computer and the calculation of PEs 
\begin{align}
	S_q = \frac{1}{1-q} \log_2 \sum_{{\vec{\sigma}}} |\langle {\vec{\sigma}} |\psi \rangle|^{2q},
	\label{sqdefSUP2}
\end{align}
is straightforward (but still requires to sum the number of terms that is exponentially large in system size $L$).
In turn, in atomic and solid state platforms ~\cite{Brydges19,Leseleuc19,Chiaro20,Pagano2020quantum,Scholl20,Ebadi20,Zeiher17,Veit21} as well as in superconducting quantum processors setups \cite{Arute19, Wu21} one can measure local observables (e.g. spin-$z$ components $Z_i$) which leads to a collapse of the state $\ket{ \psi }$ of the system on one of the basis states $\ket{\sigma}$ with probability $|\langle \sigma |\psi \rangle|^2$. Multiple repetitions of such a set of measurements on the state $\ket{\psi}$ allow to estimate the probabilities $|\langle \sigma |\psi \rangle|^2$ for all basis states $ \ket{\sigma}$ and to calculate the PEs with the formula \eqref{sqdefSUP2}. Notably, this procedure of sampling of the final state of quantum circuit $\ket{ \psi }$ was employed in the so called cross-entropy benchmarking in experiments that demonstrated quantum advantage \cite{Arute19, Wu21}. In those experiments samples of over $10^7$ output states $\ket{\sigma}$ were collected, giving us a rough estimate of the number of samples that could be collected in a hypothetical experiment that aims to observe the change of the $c_q$ term across MIPT. In the following, we describe how the PEs \eqref{sqdefSUP2} can be accurately estimated in such an experiment with superconducting quantum processor. 

Our simulation of the hypothetical experiment consists of the following steps:
\begin{enumerate}
    \item a quantum circuit is prepared in a fixed product state;
    \item quantum circuit consisting of $t=L$ unitary and measurement layers acts on the state (we employ the Haar circuit for $d=2$ studied in the Main Text, however, other circuits, consisting of gates more straightforwardly accessible in superconducting quantum processor setups \cite{Arute19, Wu21} can be considered);
    \item spin operators $Z_i$ are measured in the final state $\ket{\psi_{\mathrm{\textbf{m}}}}$ (where $\textbf{m}$ specifies the circuit realization) collapsing the state to basis state $\ket{ \sigma }$;
    \item the steps 1.-3. are repeated multiple times, and the basis states $\ket{ \sigma }$ are collected for each final state $\ket{\psi_{\mathrm{\textbf{m}}}}$.
\end{enumerate}
In our simulation the $2$-qubit unitary gates are fixed for each set of local measurements encoded in the index $\textbf{m}$ (an experimentally more convenient route would be to consider a fixed set of $2$-qubit gates). Each measurement layer consists, on average, of $p L$ local spin measurements, giving rise to approximately $2^{p L t}$ different possible states $\ket{\psi_{\mathrm{\textbf{m}}}}$ at the output of the circuit. Hence, after the first run of the experiment (steps 1.-3.), the probability to end up with the same state $\ket{\psi_{\mathrm{\textbf{m}}}}$ is $2^{-p L t}$. To gather information about the state $\ket{\psi_{\mathrm{\textbf{m}}}}$ it is necessary to perform multiple destructive measurements on it. This exponential in system size barrier in preparation a given state $\ket{\psi_{\mathrm{\textbf{m}}}}$ is the so-called problem of post-selection. The problem is common for all experimental setups that aim to observe MIPT (one possible solution is to restrict the class of circuits to circuits with space-time duality \cite{ippoliti2021postselectionfree}). As we show below, the number of measurement outcomes $\ket{ \sigma }$ for each state $\ket{\psi_{\mathrm{\textbf{m}}}}$ required to accurately estimate PEs scales exponentially with system size $L$. This scaling, combined with the fact that the number of possible final states  $\ket{\psi_{\mathrm{\textbf{m}}}}$ also scales exponentially with $L$ results in the overall exponential scaling with system size of the number of repetitions $n_s$ of the  experiment.

\begin{figure}
    \centering
    \includegraphics[width=1\columnwidth]{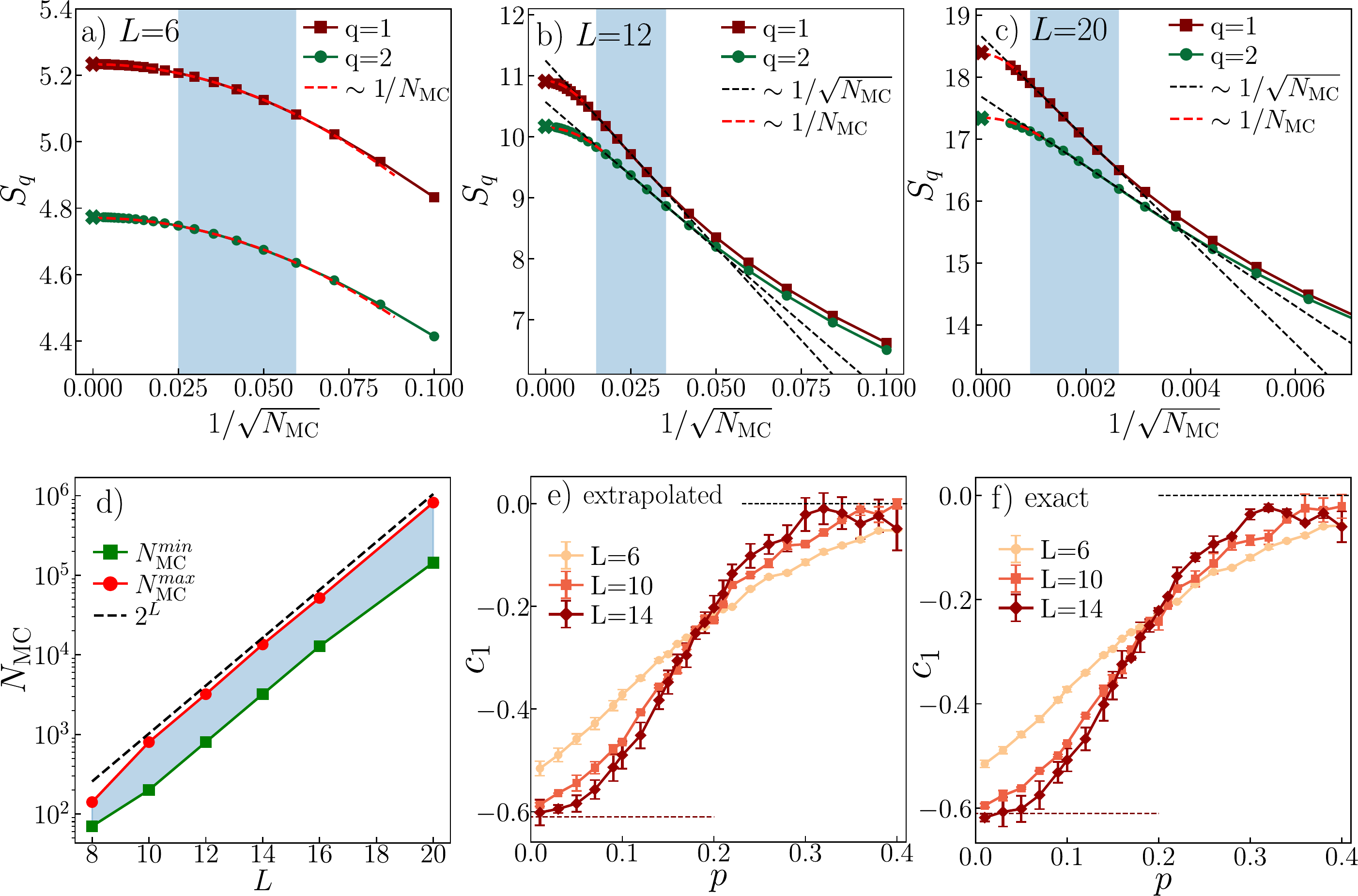}
    \caption{Simulation of an experimental observation of MIPT wit PEs on a quantum processor. Panels a), b) and c) show the estimate of PEs \eqref{sqdefSUP3} (for $q=1,2$) obtained from $N_{\mathrm{MC}}$ wavefunction snapshots. Dashed lines denote fits $g(N_{\mathrm{MC}}) \sim 1/\sqrt{N_{\mathrm{MC}}}$ (black color, using the data from the shaded area) and the extrapolating functions $f(N_{\mathrm{MC}}) \sim1/N_{\mathrm{MC}}$ (red color); the crosses at $1/\sqrt{N_{\mathrm{MC}}}=0$ denote the exact value of $S_q$; data respectively for $L=6, 12, 20$ in panels a), b), c). Panel d) shows the lower $N^{min}_{\mathrm{MC}}$ and upper $N^{max}_{\mathrm{MC}}$ bound on $N_{\mathrm{MC}}$ used in the fitting procedure, the Hilbert space dimension $\mathcal N = 2^L$ is shown for reference.
    Panel e) show the values of the sub-leading term $c_1$ determined from PEs obtained in the extrapolation scheme. Panel f) shows, for comparison, the exact value $c_1$ obtained when PEs are calculated with \eqref{sqdefSUP2}.
    }
    \label{fig:EXP}
\end{figure}
We now assume that by the measurements in step 3. we gathered, for each state $\ket{\psi_{\mathrm{\textbf{m}}}}$, in total $N_{\mathrm{MC}}$ basis states $\ket{ \sigma }$ ("wavefunction snapshots"), distributed in such a way that state $\ket{ \sigma }$ occurs $i_{\sigma}$ times in the gathered output (note that this procedure is actually a Monte Carlo sampling of the wave-function, hence the subscript "MC"). Then, the probability $|\langle \sigma |\psi \rangle|^2\approx i_{\sigma}/ N_{MC}$, and the PEs \eqref{sqdefSUP2} can be approximated as
\begin{align}
	S_q(N_{\mathrm{MC}})= \frac{1}{1-q} \log_2 \sum_{{\vec{\sigma}}} (i_{\sigma}/ N_{MC})^{q}.
	\label{sqdefSUP3}
\end{align}
The approximations of $S_q$ from our numerical experiment, averaged over $10000(100)$ final states $\ket{\psi_{\mathrm{\textbf{m}}}}$ for $L=6,12$ ($L=20$) are shown as a function of the number of snapshots $N_{\mathrm{MC}}$ in Fig.~\ref{fig:EXP}~a),~b),~c). We observe that $S_q(N_{\mathrm{MC}})$ is an increasing function of $N_{\mathrm{MC}}$ and that the increase persists even for $N_{\mathrm{MC}}$ significantly larger than the Hilbert space dimension $\mathcal N=2^L$. However, the functional dependence $S_q$ on $N_{\mathrm{MC}}$ is rather simple. For small system sizes ($L=4,6$, see Fig.~\ref{fig:EXP}~a)) we observe that $S_q(N_{\mathrm{MC}}) \sim 1/N_{\mathrm{MC}}$. Fitting a first order polynomial in $1/N_{\mathrm{MC}}$ in interval $N_{\mathrm{MC}} \in[ 250,1500 ] $ and extrapolating it to $ N_{\mathrm{MC}}\rightarrow \infty $ allows us to reproduce the exact value of $S_q$ (calculated with the full knowledge of many-body wave-function with formula \eqref{sqdefSUP2}) with accuracy better than $0.1\%$. 
For larger system sizes ($L \geq 8$) we observe that $S_q(N_{\mathrm{MC}}) \sim 1/\sqrt{N_{\mathrm{MC}}} $ in the interval from $N_{\mathrm{MC}} \approx \mathcal N / 8$ to $N_{\mathrm{MC}} \approx \mathcal N $ and that for larger $N_{\mathrm{MC}}$ we again have $S_q(N_{\mathrm{MC}}) \sim 1/N_{\mathrm{MC}}$. This observation motivates the following extrapolation scheme.
\begin{itemize}
    \item The curve $S_q(N_{\mathrm{MC}})$ is fitted by a function $f(N_{\mathrm{MC}}) = a/\sqrt{N_{\mathrm{MC}}}+b$ (where $a,b$ are fit parameters) in an interval $N_{\mathrm{MC}} \in [N^{min}_{\mathrm{MC}}, N^{max}_{\mathrm{MC}}]$.
    \item For $N_{\mathrm{MC}}>\alpha_q \mathcal N$ we define a function $g(N_{\mathrm{MC}}) = a_0 \left(1/N_{\mathrm{MC}} - 1/(\alpha_q \mathcal{N}) \right) + f( \alpha_q \mathcal N)$, where $a_0$ is chosen in such a way that $\frac{dg(N)}{dN}|_{N = \alpha_q \mathcal N} $ equal to $\frac{df(N)}{dN}|_{N = \alpha_q \mathcal N} $; $\alpha_q$ is fixed for all system sizes for given $q$ (we take $\alpha_1=2$ and $\alpha_2=0.7$).
    \item The desired value of PE is obtained by the extrapolation: $ S_q = \lim_{N_{\mathrm{MC}} \rightarrow\infty } g(N_{\mathrm{MC}})$.
\end{itemize}
The extrapolation scheme is illustrated in Fig.~\ref{fig:EXP}~b),~c) where the fits and the functions $f(N_{\mathrm{MC}})$ and $g(N_{\mathrm{MC}})$ are denoted by dashed lines. The scheme allows to estimate the value of PEs with accuracy of about $0.1\%$ for all system sizes considered. The experimental resources needed for the determination of $S_q$ are proportional to $N^{max}_{\mathrm{MC}}$ which scales exponentially with $L$ as visible in Fig.~\ref{fig:EXP} but remains bounded from above by the Hilbert space dimension $\mathcal N = 2^L$. A rough estimate of the total number of repetitions $n_s$ of steps 1.-3. can be obtained as follows: there is approximately $\approx 2^{pLt}$ possible final states $\ket{\psi_{\mathrm{\textbf{m}}}}$ (assuming there is no randomness in the 2-qubit gates), for each of them one needs to perform at least $\approx N^{max}_{\mathrm{MC}} \approx 2^L$ measurements. This gives $n_s \sim 2^{pLt + L}$. For instance, assuming that for system size $L=10$ and for measurement probability $p=0.1$ the circuit of depth $t=10$ is sufficient to reach the saturation value of $S_q$, we get $n_s = 2^{20} \approx 10^6$. For larger values of $p$, the steady state value of $S_q$ is reached for a circuit of lower depth (for growing $p$ the value of $t$ needed to reach the steady state decreases) and the number of samples $n_s \approx 10^6$ seems to be a reasonable estimate for any value of $p$. Such a number of samples can be gathered easily with superconducting quantum processors (more than $10^7$ samples or output were gathered in \cite{Wu21} for much larger system size). This should be compared with systems of up to $L=8$ sites considered in the recent experiment \cite{noel2021observation}. However, $n_s$ increases very quickly with $L$: an analogous estimate for $p=0.1$ and $L=14$ gives $n_s \approx 10^{10}$. 
If the problem of post-selection was eliminated (note that this problem does not occur at $p\approx 0$), system sizes beyond $L=20$ could be reached with present day superconducting quantum processors.

\end {document}